\documentclass[twocolumn,amsmath,floatfix,prb,aps,superscriptaddress]{revtex4}

\usepackage{graphicx}
\usepackage{epsfig} 
\usepackage{dcolumn}
\usepackage{bm}
\usepackage{times}

\makeatletter

\begin{document} 

\title{ 
Kondo-Anderson transitions }

\author{S. Kettemann} 

\affiliation{Jacobs University, School of Engineering and Science,
  Campus Ring 1, 28759 Bremen, Germany  and Division of Advanced
  Materials Science Pohang University of Science and Technology
  (POSTECH) San 31, Hyoja-dong, Nam-gu, Pohang 790-784, South Korea}

\author{E. R. Mucciolo} 

\affiliation{Department of Physics, University of Central Florida,
  P.O. Box 162385, Orlando, Florida 32816, USA}

\author{I. Varga} 

\affiliation{Elm\'eleti Fizika Tansz\'ek, Budapesti M\H uszaki \'es
  Gazdas\'agtudom\'anyi Egyetem, H-1521 Budapest, Hungary}

\author{K. Slevin} 

\affiliation{Department of Physics, Graduate School of Science, Osaka
  University, 1-1 Machikaneyama, Toyonaka, Osaka 560-0043, Japan}

\date{\today}

\begin{abstract} 
Dilute magnetic impurities in a disordered Fermi liquid are considered
close to the Anderson metal-insulator transition (AMIT).
 Critical Power law correlations
   between electron wave functions at different energies in the vicinity of the AMIT
    result in the formation of 
    {\it pseudogaps}  of the local density of states.
  Magnetic impurities can remain unscreened at such sites. We 
     determine 
the density of the resulting free magnetic moments in the zero temperature limit.
While it is finite on the insulating side of the AMIT, it
  vanishes at the AMIT, and decays 
   with a power law as function of  the distance to  the AMIT. 
   Since the fluctuating spins of these free magnetic moments break the time
    reversal symmetry of the  conduction electrons, we find a shift of the AMIT, 
    and the appearance of a semimetal phase.    
  The distribution function of the Kondo temperature $T_{K}$ is derived 
  at the AMIT, in the metallic phase and in the insulator phase.
   This allows us  to find the quantum phase diagram in 
    an external magnetic field $B$ and at  finite temperature $T$.
We calculate the resulting magnetic susceptibility, the
specific heat, and the spin relaxation rate as function of
temperature. We find a phase diagram with finite temperature
transitions between insulator, critical semimetal, and metal
phases. These new types of phase transitions are caused by the interplay between
Kondo screening and Anderson localization, with the latter being
shifted by the appearance of the temperature-dependent spin-flip
scattering rate. Accordingly, we  name them
Kondo-Anderson transitions (KATs).
\end{abstract}



\maketitle

\section{Introduction}

The non-Fermi liquid behavior of disordered electronic systems such
as the power-law divergence of the low-temperature magnetic
susceptibility, can originate from a wide distribution of the Kondo
temperatures of magnetic
impurities.\cite{bhatt,langenfeld,dobros,Miranda} The Kondo
temperature $T_{K}$ is exponentially dependent on the local exchange
coupling $J$ and thus on the hybridization of a magnetic impurity
state with the conduction band. Since the hybridization is
proportional to the overlap integral between a localized magnetic
impurity orbital and the conduction band, it can be exponentially
sensitive to microscopic variations of the position of a magnetic
impurity.\cite{bhatt} The exchange coupling also depends on all wave
function amplitudes of the conduction electrons at that position, and
thus on the local density of states (LDOS) in the conduction band. The
distribution of the LDOS depends strongly on the nonmagnetic disorder
strength and is known to attain a wide log-normal
distribution.\cite{lernerldos}  In the earliest
approaches to this problem it was argued that 
the LDOS in the vicinity of  the Fermi
energy depends only slowly on energy. Therefore,  the small $T_{K}$
 tail of the distribution  should be directly connected to the distribution of the LDOS at
  the Fermi energy, which results in a wide distribution of $T_{K}$ on the metallic side
   of the AMIT whose width increases as the AMIT is approached.\cite{dobros} 
 However, since $T_{K}$ is determined  by an integral
over all energies, the fluctuations of the LDOS are to some
degree averaged out so that $T_{K}$ may not vary as strongly.
 In the weak-disorder limit, explicit analytical
calculations show that the width of the distribution of Kondo
temperatures is finite due to {\it correlations between wave functions
  at different energies}. In disordered metals, these correlations are
induced by the diffusion of conduction electrons. However, the
resulting width was found to be small in three dimensions in the
weak-disorder limit.\cite{kettemannjetp,micklitz} For strong disorder
 the distribution of
$T_{K}$ has been studied by means of the exact numerical diagonalization
of finite systems to be wide and bimodal.\cite{imre,cornaglia}  
 Its
low-$T_{K}$ power-law tail has been argued to have a
 universal power.\cite{cornaglia}  Similar bimodal distributions 
 of $T_{K}$ have been found in 2D disordered metals using the nonperturbative numerical 
 renormalisation group and the quantum Monte Carlo 
  methods.\cite{zhuravlev}
 
In this paper, we consider dilute magnetic impurities in a disordered
Fermi liquid close to the Anderson metal-insulator transition
(AMIT). We use information about the multifractality of the critical
wave functions at and in the vicinity of the AMIT,
\cite{multifractal,rmpmirlinevers,rodriguez} in order to derive
physical properties arising from the interaction between the
conduction electron spins and the quantum spins of the magnetic
impurities at the transition point. In particular, we obtain
analytical expressions for the distribution functions of the
Kondo temperature $T_{K}$ at and in the vicinity of the AMIT. 

 At finite density, the magnetic moments become coupled by 
 the indirect exchange coupling 
$J^{I}$. \cite{dobros} The latter can be  calculated  using the expression for the 
  generalized RKKY
coupling,\cite{RKKY} which is a function of  the local
exchange coupling $J$.\cite{liechtenstein} 
   The  distribution of $J^{I}$ and its competition with the Kondo screening will be studied in a 
    subsequent article. \cite{rkkykondo} Here, we consider the dilute limit when the coupling between 
     the moments can still be disregarded. 
In the
metallic phase the density of free, unscreened  magnetic moments $n_{FM}$ is found
to vanish exactly in the zero-temperature limit. In the insulating
phase, $n_{FM} (T=0)$ is finite and increases with a  power law
 as function of the  distance to the AMIT.

In Sec. \ref{sec:multfrac} we begin with 
 a brief review of multifractal statistics.
Wave functions in the vicinity of the AMIT are furthermore power-law correlated in
energy.\cite{powerlaw,cuevas,ioffe,ioffe2} As has been noted in
Ref. \onlinecite{cuevas} this has the surprising consequence that
multifractal eigenfunctions which are close in energy are likely to
have their maximal intensity at the same positions in space. This
positive correlation has been called stratification\cite{cuevas} and
is opposite to what is expected in the localized phase where states
close in energy have their maximal intensity most likely at far apart
positions in space.
Another consequence is the opening of local pseudogaps at positions
where the critical wave functions have vanishing
intensity.\cite{kettemann} At other sites the
local intensity diverges with a power law. 

 In Sec. \ref{sec:Kondo}
the Kondo problem 
 in disordered systems is formulated.

  In
Sec. \ref{sec:Kondoamit}
the
distribution of Kondo temperatures at the AMIT is derived and discussed. 

 In Sec. \ref{sec:metallic} we extend the analysis to the
metallic regime where the Fermi energy is located above the mobility
edge. Although all states in the vicinity of the Fermi energy are
extended, the multifractal nature of the eigenstates on length scales
smaller than the correlation length $\xi$ leads to fluctuations of the
LDOS and modifies the magnetic properties of the system as we review in Sec. \ref{sec:metallic}. 

In
Sec. \ref{sec:insulator} we extend the analysis to the insulating
regime when the Fermi energy is located below the mobility edge and
the wave functions are localized exponentially with a localization
length $\xi_{c}$. Multifractal fluctuations still occur on length
scales smaller than $\xi_{c}$ and the wave functions are power law
correlated within a localization volume. It is therefore crucial to
take these effects into account in order to get the correct
distribution of Kondo temperatures  in
the insulating phase . 

In Sec. \ref{sec:insulator2D} we summarise the results for the 
 2D system, where all states are localised. 

In Sec. \ref{sec:quantumphase} the quantum
phase diagram of the Anderson metal-insulator transition in the
presence of magnetic impurities is derived and plotted as function of
the exchange coupling parameter $J$ and disorder amplitude $W$. We
establish the existence of a {\it critical semimetal phase} where both
the correlation and the localization lengths are infinite within a
finite interval of disorder amplitudes, and the conductivity is
vanishing. 

In Sec. \ref{sec:magneticfield} we consider how the quantum phase diagram 
 changes in an external magnetic field, which couples via the 
  Zeeman term to the magnetic impurities and via the orbital term to the 
   conduction electrons. 
   
   In Sec. \ref{sec:nfl}
    we present the results for the  Non-Femi-liquid properties, 
 in particular  the magnetic susceptibility, the specific heat, and the
spin relaxation  rate as functions of temperature, concentration of magnetic
moments, and disorder amplitude.

In Sec. \ref{sec:finiteT} we study the consequences of the
temperature dependence of the spin relaxation rate, which is caused by
the Kondo effect. This leads to transitions between insulator,
critical semimetal, and metal phases at finite temperatures. These one
may call, accordingly, {\it Kondo-Anderson transitions}.

 In
Sec. \ref{sec:conclusions} we provide our conclusions, comment on
experimental realizations of these transitions, and discuss remaining
open problems.

 In Appendix \ref{sec:A} we review the wave function
correlations in the vicinity of the AMIT when one state is at the
mobility edge. The joint distribution function of eigenfunction
intensities is derived such that it matches the critical correlation
functions. Next, the correlation  function and the joint
distribution functions are derived when both eigenstates are
located away from the mobility edge of the AMIT.
 We also discuss and present results on higher moment correlation functions. 
 
  In Appendix \ref{sec:B} we derive the function 
   $F(\alpha,T_{K})$ as defined in Eq. (\ref{tkc3}).

\section{ Multifractality, Local Pseudogaps and Power Law Divergencies}
\label{sec:multfrac}
 
  The AMIT is known to be a 2nd order quantum phase transition, 
   where both the localization length and the correlation length diverge
    to infinity. At the AMIT the electrons are in a critical state, 
     which is neither extended nor localized, but sparse, as seen  
  in Fig. \ref{fig:mf} where the intensity $|\psi_l({\bf r})|^2$ at the AMIT is plotted. 
  
   \subsection{Multifractality}
\begin{figure}[t]
\vspace{-2cm}
\includegraphics[width=10.5cm]{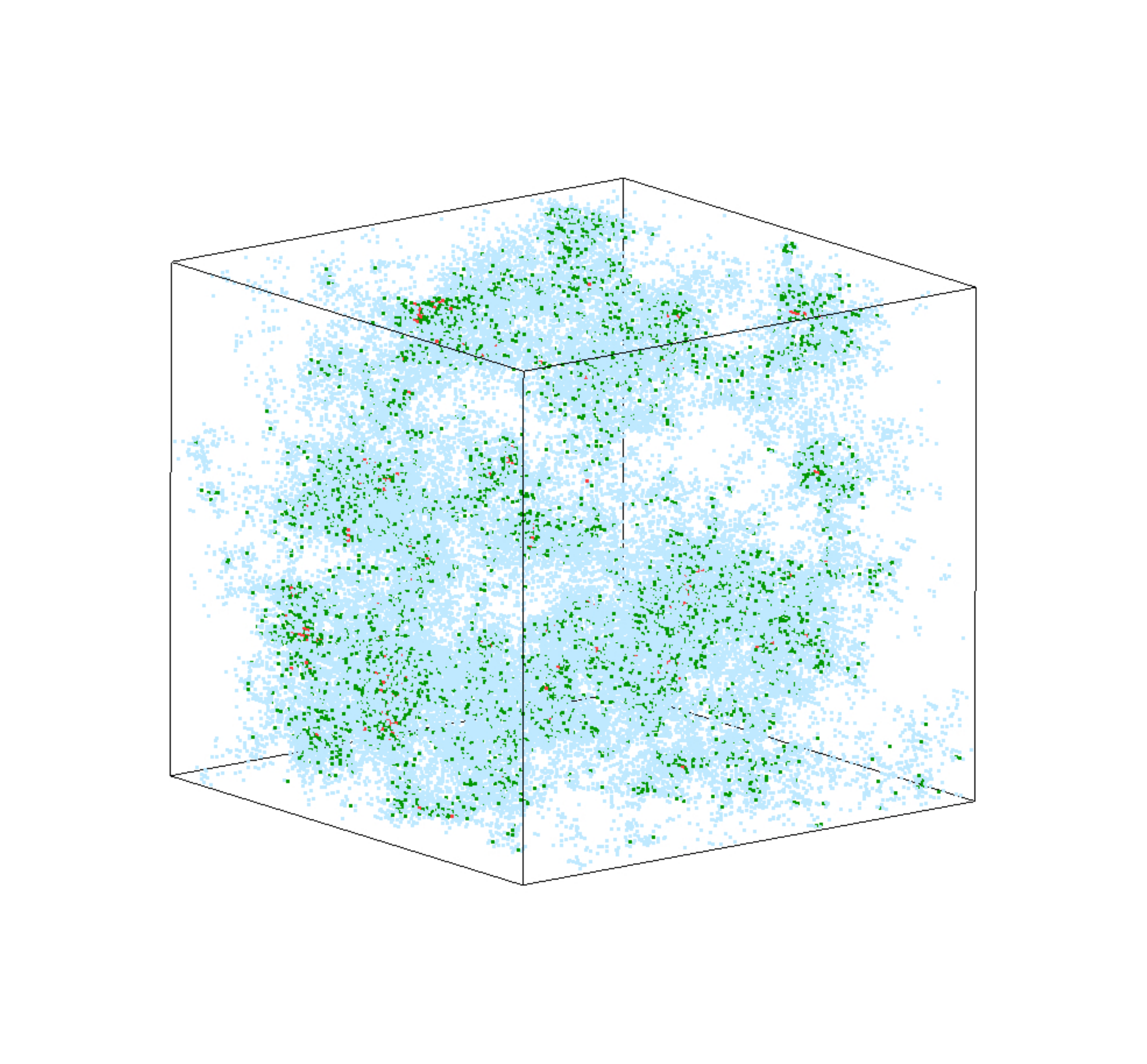}
\vspace{-2cm}
\caption{(Color online) The
 local  intensity is plotted here  for a
  critical state at the three-dimensional AMIT, as obtained by exact diagonalisation of
   the Anderson tight binding model with $W=16.5 t$, with hopping parameter 
    $t$ at energy $E=2t$ on a cubic lattice with spacing $a$, and
   size $L=100 a$.  The coloring of the plotted intensity  was done 
    according to   
    red: $\alpha \in [ 1.2,1.8]$, green: $ \alpha \in [1.8,2.4] $, light blue: $\alpha \in [2.4,3.0]$, where 
     is defined by 
     $ \alpha = -\ln |\psi|^2 / \ln L$. Sites with higher intensity, $\alpha < 1.2$,
      are so rare that their occurrence can not be resolved in this plot. All other
       sites whose intensities are not plotted correspond to 
        lower intensity   $ \alpha > 3.0$. 
Thereby,   about 80 $\%$ of the total state intensity is shown.}
\label{fig:mf}
\end{figure}

  This critical state can be characterized by 
 the moments of eigenfunction intensities $|\psi_l({\bf r})|^2$, which 
scale as powers of the system linear size $L$,
\begin{equation}
\label{single}
P_q = L^d \langle\, |\psi_l({\bf r})|^{2 q} \rangle \sim L^{-\tau_q},
\end{equation}
where $d$ is the spatial dimension. 
  In a metal   the powers $\tau_q$   would be  given by $d (q-1)$.
   Critical states are characterized by 
    multifractal dimensions $d_{q} <d$ which may change with  the power $q$ of the moments. 
     These are  related to the exponents of the q-th moments by  $\tau_q = d_{q} (q-1) $.
The corresponding distribution function of the intensity
is close to  log-normal in  good
approximation,\cite{rmpmirlinevers}
\begin{equation}
\label{eq:Pone}
P(| \psi_l({\bf r}) |^2) = \frac{1}{| \psi_l({\bf r}) |^2} 
L^{ - \frac{(\alpha-\alpha_0)^2}{2   \eta  } },
\end{equation}
where $ \alpha = - \ln | \psi_l({\bf r}) |^2 / \ln L $, $  \eta  = 2 (\alpha_{0}-d)$ and 
$\alpha_0>d$.   The multifractal dimension $d_{q}$ is then 
 related to $\alpha_{0}$ by $d_{q}= d - q (\alpha_{0}-d) $
  for not too large $q$.  
  At $q_{c}= \alpha_{0}/2/(\alpha_{0}-d)$ 
   there is a termination of $\tau_{q}$  so that 
    it remains constant $ \tau_{q} = \tau_{q_{c}}$ for $q>q_{c}$.\cite{rmpmirlinevers}
  Throughout this article we assume the validity 
   of this Gaussian  distribution of $\alpha$. 
    In Fig. \ref{fig:mf} the
 local  intensity is plotted   for a
  critical state at the three-dimensional AMIT as obtained by exact diagonalisation of
   the Anderson tight binding model with a box distribution of width 
   $W=16.5 t$. Here, $t$ is  the hopping parameter. The energy is 
 at  $E=2t$ on a cubic lattice with spacing $a$, and  linear lattice size 
   size $L=100 a$.  The coloring of the plotted intensity  was done 
    according to the amplitude of   $ \alpha = -\ln |\psi|^2 / \ln L$.
     Sites with higher intensity with $\alpha < 1.2$
      are so rare that their occurrence can not be resolved in this plot. All other
       sites whose intensities are not plotted correspond to 
        lower intensity with  $ \alpha > 3.0$. 
Thereby,   about 80 $\%$ of the total state intensity is shown  in Fig. \ref{fig:mf}.

 \subsection{Critical Correlations}
    The wave function intensity
of a state at energy $E_{l}$ 
at a certain coordinate   ${\bf r}$ has a power-law correlation with the
intensity at the mobility edge energy $E_{M}$ with the power $  \eta/d $,
 which is  related to $\alpha_{0}$ by
\begin{equation}
  \eta  = 2(\alpha_0-d),
\end{equation}
(see Eq. (\ref{criticalcorrelations}) in  Appendix A).\cite{powerlaw,cuevas} 
As noted in Ref. \onlinecite{mirlin} this correlation is due to the fact
 that eigenstates in the vicinity of the AMIT have a similar  multifractal envelope function
 $\phi_{l}({\bf r})$ and can be written as $\psi_l ({\bf r})= \phi_l ({\bf r}) \chi_{ul}  ({\bf r})$.
  Here
$ \chi_{ul}  ({\bf r})$  denotes a factor of the  wave function which fluctuates independently from 
$\phi_{l}({\bf r})$ on short, microscopic length scales. 

{\it Joint Distribution Function.}
One can take this correlation  into account 
 by considering the joint distribution function of two wave functions. 
 In order to obtain the correlation function  Eq. (\ref{criticalcorrelations})
  and the distribution function of a single state, Eq. (\ref{eq:Pone}), it should 
 be of the form
  \begin{equation}
\label{eq:Ptwo}
P(\alpha_l,\alpha_k) = \xi_{l}^{ a_{l k} \left[ f (\alpha_l) -d
    \right]} L^{ a_{l k} \left[ f (\alpha_k) -d \right]} K_{l
  k}^{- a_{l k} \frac{(\alpha_l - \alpha_0)(\alpha_k -\alpha_0) }{d
      \eta }},
\end{equation}
 where we obtained  $K_{{lk}} = {\rm Max} \left[ | E_{l} -E_{k} |,
  \Delta_{\xi_{l}} \right]/E_{c} $, see Appendix A for the derivation 
   and definitions of $a_{lk}$. Here, $\xi_{l}$ is the correlation length/ localization length 
    of  the state with energy $E_{l}$ on the metallic/insulating side of the transition,
     and $\Delta_{\xi_{l}}= D/\xi_{l}^d$. 
  
  {\it Conditional Intensity.} 
   Thus,  we can derive the conditional
intensity of a state at energy $E_{l}$ given that  the intensity at
the critical energy $E_{k}=E_{M}$ is $|\psi_M ({\bf r})|^2 =
L^{-\alpha}$. This conditional intensity, relative to the intensity of
that of an extended state, is  obtained by averaging over the 
 joint distribution function Eq. (\ref{eq:Ptwo}) (for the derivation see Appendix A), \cite{kettemann}
\begin{eqnarray}
\label{cci}
I_{\alpha} & = & L^d \langle| \psi_{l} ({\bf r}) |^2
\rangle_{|\psi_M({\bf r})|^2=L^{-\alpha}} \nonumber \\ & = & \left|
\frac{E_{l} -E_{M} }{E_{c}} \right|^{r_{\alpha}},
\end{eqnarray}
where the power is given by
\begin{equation} \label{power}
r_{\alpha}=\frac{\alpha-\alpha_{0}}{d} -
  \frac{  \eta }{2 d } g_{lM}.
  \end{equation}
This result is valid for  $| E_{l} -E_{M} | < E_{c}$, where $E_{c}$ is the energy scale over
which the critical correlations exist and  $g_{lM} = \ln | (E_{l}
-E_{M}) /E_{c} |/ (d \ln L) $. The average is 
 done over the intensity $| \psi_{l} ({\bf r}) |^2$
  using the conditional distribution function, Eq. (\ref{conditional}).
   fixing the intensity $L^{-\alpha}$ at the AMIT.
  Typically, $E_c$ is a fraction
of the bandwidth $D$: $E_{c} \sim D/(2d \ln 2d)$.\cite{powerlaw} When $E_{l}$ is located at a finite energy interval away from
the mobility edge $E_{M}$, the coefficient $g_{lM} = \ln | (E_{l}
-E_{M}) /E_{c} |/ (d \ln L) $ vanishes for $L \rightarrow \infty$. 
Close to $E_{M}$ the coefficient saturates: $ g_{lM}
|_{E_{l}\rightarrow E_{M}} \rightarrow -1 $ and Eq. (\ref{cci})
reduces to $L^{d-\alpha}$, the local intensity at $E_{M}$ relative to
the intensity of an extended state $L^{-d}$. Note that in Eq. (\ref{cci}) the average over the 
uncorrelated factor of the 
 wave function 
$ \chi_{u l}  ({\bf r})$ has been done, which is $\langle | \chi_{u l}  ({\bf r})|^2 \rangle=1$.
         
\begin{figure}[t]
\includegraphics[width=8cm,angle=0]{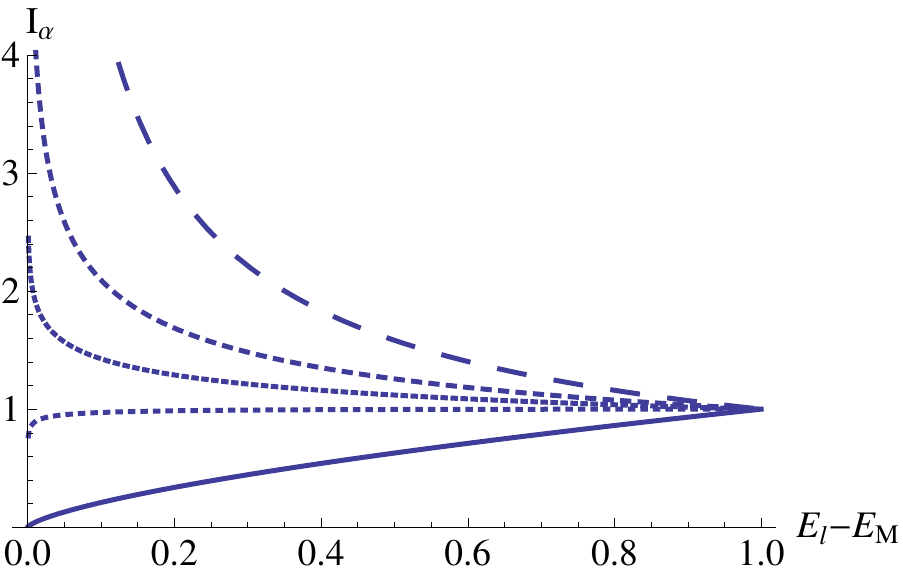}
\caption{(Color online) The conditional intensity $I_{\alpha}$ relative to the intensity of an
  extended state as function of the distance in energy to the mobility
  edge in units of the correlation energy $E_{c}$. The exponent
  $\alpha$, which is related to the intensity $L^{-\alpha}$ at
  $E_{M}$, takes the values $\alpha = 2,3,3.5,4,6$ in the sequence of
  decreasing dashing. Local pseudo gaps are seen for $\alpha >
  \alpha_{0}$, while local power law divergences occur for $\alpha <
  \alpha_{0}$. Here, we set $d=3$ and $\alpha_{0}=4$.}
\label{fig:pseudogap}
\end{figure}

{\it  Local Pseudogaps.} From Eq. (\ref{cci}) we see that at positions
in space where the local wave function intensity at the mobility edge
is small, corresponding to  $\alpha $  larger than its typical value 
$\alpha_{0}$, the wave function intensity
is suppressed within an energy range of order $E_{c}$ around $E_{M}$.
  Thereby  local pseudogaps are formed with a power $r_{\alpha} =
\frac{\alpha-\alpha_{0}}{d}$ with vanishing LDOS at the 
 mobility edge, as shown in Fig. \ref{fig:pseudogap}.

{\it Local Power Law Divergency.} On the other hand, when the
intensity at the mobility edge is larger than its typical value
$L^{-\alpha_{0}}$, which corresponds to $\alpha < \alpha_{0}$, the
LDOS is enhanced within  an energy range of order $E_{c}$ around
$E_{M}$, increasing as a power law when $E_{l}$ approaches the
mobility edge, as shown in Fig. \ref{fig:pseudogap}.

\section{Kondo Effect in a Disordered Electron System }
\label{sec:Kondo}
\subsection{Kondo Impurity Hamiltonian}
Magnetic impurities are generally described by an Anderson model where
a localized level with energy $\epsilon_d$ and on-site Coulomb
repulsion $U$ hybridizes with electrons in the conduction band which
is described by a Hamiltonian $H_0$.\cite{andersonmm} This Hamiltonian
may include the random potential of nonmagnetic impurities $V({\bf
  r})$. Using the eigenstates $\{\psi_n\}$ and energies $\{E_n\}$ of
$H_0$, with the corresponding one-particle density operator given by
$\hat{n}_{n,\sigma}$, the Anderson Hamiltonian is written as
\begin{eqnarray}
H_A &=& \sum_{n, \sigma} E_n\, \hat{n}_{n,\sigma} + \epsilon_d
\sum_{\sigma} \hat{n}_{d,\sigma} + U\, \hat{n}_{d,+} \hat{n}_{d,-}
\nonumber \\ & & +\ \sum_{n,\sigma} \left( t_{nd}\, c_{n \sigma}^+
d_{\sigma} + \mbox{H.c.} \right),
\end{eqnarray}
where $\hat{n}_{d,\sigma} $ is the density operator of the impurity
level. The hybridization amplitude $t_{nd}$ is proportional to the
eigenfunction amplitude at the position of the magnetic impurity
$\psi_n^\ast(0)$ and to the localized orbital amplitude $\phi_d (0)$:
$t_{nd} = t\, \psi_n^* (0) \phi_d (0)$. One can employ the
Schrieffer-Wolff transformation,\cite{hewson,pseudogap} formulated in
terms of eigenstates $\psi_n$ to take into account double occupancy up
to second order in $t_{nd}$.\cite{kettemannjetp} The result is an {\sl
  s}-{\sl d} contact Hamiltonian with exchange couplings given by
\begin{equation}
\label{jnn}
J_{n n^\prime} = t_{n d}^\ast\, t_{n^\prime d} \left( \frac{1}{U +
  \epsilon_d - E_{n^\prime}} + \frac{1}{E_n - \epsilon_d} \right)
\end{equation}
and an additional potential scattering term with amplitude
\begin{equation}
\label{knn}
K_{n n^\prime} = t_{n d}^\ast\, t_{n^\prime d} \left( - \frac{1}{U +
  \epsilon_d - E_{n^\prime}} + \frac{1}{E_n - \epsilon_d} \right).
\end{equation}

Note that in the symmetric Anderson model, $K_{n n^\prime}$ vanishes
for all $n,n^\prime$ when $|E_{n,n^\prime} - E_F| \ll U$, since, in
this case $\epsilon_d = E_F- U/2$, where $E_F$ is the Fermi
energy. For arbitrary $\epsilon_d$ wave functions with small
amplitude at the position of the magnetic impurity are hardly modified
by this potential scattering term since $K_{n n^\prime} \sim \psi_n^*
(0) \psi_{n^\prime}(0) $. Hence, we will first retain only the
exchange couplings $J_{n n^\prime} = J \psi_n^\ast (0)
\psi_{n^\prime}(0)$, with $J \sim t^2/U$, and leave the discussion of 
possible  effects of finite $K_{n,n'}$ to Sec. \ref{sec:conclusions}.

\subsection{Kondo Temperature}

As we have shown previously with the numerical renormalization group
and the quantum Monte Carlo methods,\cite{zhuravlev} the distribution
of Kondo temperatures is in very good qualitative agreement with the
one obtained from the one-loop equation of Nagaoka and
Suhl.\cite{suhl}  Only,  its average needs to be rescaled, in order to account for a
shift of the distribution towards larger values of $T_{K}$ due to the
higher-loop corrections.\cite{zu96,zhuravlev} Therefore, we will use here
the one-loop equation to calculate $T_{K}$, as given by
\begin{equation}
\label{tkc}
1 = j \frac{\Delta}{2} \sum_{l } \frac{
L^d | \psi_{l} ({\bf r}) |^2
 }{   \epsilon_l}
\tanh \left(\frac{\epsilon_l}{2 T_K} \right) \equiv F(\{\alpha_{l}\},T_{K}),
\end{equation} 
where  $j=J/D$, $\Delta = D/N$ is the mean level spacing and 
 $\epsilon_{l}=E_{l}-E_{n}$. $N$ is the number of states in the sample of 
  volume $V=L^d$.  
   This defines the Kondo-temperature in terms of the local intensities at all energies
    $\epsilon_{l}$ in the sample. 
    
    \subsection{Distribution Function of the Kondo temperature}
    Thus, one can derive the distribution function 
     of $T_{K}$, when the distribution function of all intensities $P( \alpha_{1},...,\alpha_{l},...,\alpha_{N})$
       is known, by solving
       \begin{equation} \label{ptkexact}
       P(T_{K}) = \int \prod_{l} d \alpha_{l}  P ( \{\alpha_{l}\})
       \delta (1- F(\{\alpha_{l}\},T_{K})) |\frac{dF}{d T_{K}}|,
       \end{equation}
        where $F(\{\alpha_{l}\},T_{K})$ is defined by Eq. (\ref{tkc}).
        Note that $F(\{\alpha_{l}\},T_{K})$ is always a decreasing function 
   of $T_{K}$, so that  $ |\frac{dF}{d T_{K}}|= -  \frac{dF}{d T_{K}}$.
       
          \section{Kondo Effect at the Anderson-Metal-Insulator Transition}
          \label{sec:Kondoamit}
   {\it  Conditional Average. }       In a first 
           attempt to get its distribution function    we calculate the Kondo temperature $T_{K}
$ for a given intensity $\xi^{-\alpha}$  at the Fermi energy, and integrate over all 
 other         intensities with the conditional distribution function Eq. (\ref{conditional})
  for fixed $\alpha$.  Thereby we find   $T_{K}(\alpha)$ as a function of  $\alpha$,
using that the conditional intensity of state $l$, $\langle| \psi_{l} ({\bf r}) |^2
\rangle_{|\psi_n({\bf r})|^2=L^{-\alpha}}$  is given by Eq. (\ref{cci}), 
\begin{equation}
\label{tkc3}
1 = j \frac{\Delta
 }{  E_{c}} \sum_{\epsilon_{l} < E_{c} }
 \left|
\frac{\epsilon_{l}  }{E_{c}} \right|^{r_{\alpha}-1}\tanh \left(\frac{\epsilon_l}{2 T_K } \right) \equiv F[\alpha,T_{K}],
\end{equation} 
  where  the summation over $l$ is restricted  to energies within the energy interval 
   of  the correlation Energy
   $E_{c}$ around the mobility edge.\cite{remark}
   
We note that Eq. (\ref{tkc3}) defines the Kondo temperature in a system
with {\it pseudogaps} of power $r_{\alpha}$, Eq. (\ref{power}), in the local density of states
 shown in Fig. 2, 
   when this power is
positive.\cite{pseudogap} Therefore, the Kondo temperature is reduced at such sites,
since the magnetic impurities are placed in the locally suppressed
LDOS of the conduction electrons.  

On the other hand, at sites where
the power $r_{\alpha}$ is negative  the opposite occurs and $T_{K}$
is enhanced. The Kondo problem with a {\it power-law divergence} in
the density of states was studied in Ref. \onlinecite{bulla}, 
 finding an enhancement of $T_{K}$ towards the strong coupling fixed point
  given by $J$. Since
$\alpha$ is distributed over all values from $0$ to 
 infinity  according to Eq. (\ref{eq:Pone}), we find by solving Eq. (\ref{tkc3}) that
$T_{K} (\alpha)$ is distributed, accordingly. We note that Eq. (\ref{tkc}) is averaged over
 the uncorrelated part of the wave function, $ \chi_{u l}  ({\bf r})$. Since
  these short scale fluctuations are uncorrelated in energy, they are averaged out by the 
   summation over energy levels $l$ in Eq. (\ref{tkc3}), and 
    gives rise  to small fluctuations of  $T_{K}$ of order of $1/N$, only. 
    Therefore their effect on the distribution of 
    $T_{K}$ is negligible in the thermodynamic limit.

We can now proceed and solve Eq. (\ref{tkc3}) analytically in various
limits. The Kondo temperature for a given intensity $  | \psi_n({\bf r}) |^2 = L^{-\alpha}$ at
the Fermi energy  is for $\alpha > \alpha_{0}-d$ found to be given by
\begin{equation}
\label{tkalpha}
\frac{T_K}{E_c} = \left[ \left(1 - \frac{\alpha - \alpha_0}{d}
  \frac{1}{j} \right) c_{\alpha} \right] ^{\frac{d}{\alpha -
    \alpha_0}}.
\end{equation}
Here, $j=J/D$ and $c_{\alpha} = \frac{2 \alpha-\eta}
{\alpha-\eta/2+d} $. Eq. (\ref{tkalpha}) takes into account
the critical correlations at the AMIT.
 In deriving it (see Appendix \ref{sec:B} for details), we  approximate $\tanh x \approx x$ for $x < 1$, and $\tanh x \approx 1$ 
 for $x > 1$, yielding two terms. One includes the integral over
  all energies within a window of width $2 T_{K}$ around the Fermi energy.
   The other one is over all larger energies in the conduction bands. 
     In order that the resulting equation is valid for all $\alpha$ it is important to keep 
    both terms.
   
 For the typical value
$\alpha=\alpha_{0}$ we recover $T_K$ of a clean system in the one-loop
approximation, namely, 
\begin{equation}
T_K(\alpha = \alpha_{0}) \sim E_{c} \exp(- 1/j) \sim T_K^{(0)}.
\end{equation}
 From
Eq. (\ref{tkalpha}) we see that at particular sites where the wave
function amplitude is large, corresponding to $\alpha < \alpha_{0}$,
$T_{K}$ becomes enhanced. In particular, for a wave function intensity
comparable to the one of a metallic state, $\alpha =d$, we find that
\begin{equation}
T_{K} (\alpha =d) \sim j^{2/  \eta/d }.
\end{equation}
 This is larger than
$T_K^{(0)}$ when $j < 1$. 

\subsection{ Density of Free Magnetic Moments at Zero Temperature.} 

At sites where the wave function amplitude
is small, corresponding to $\alpha > \alpha_{0}$, $T_{K}$ is
suppressed due to the appearance of local pseudogaps.
The Kondo temperature in the presence of pseudogaps of power $r$
 is well known to vanish when the exchange coupling 
  does not exceed a critical value $j_{c} = r $.\cite{pseudogap}
   Since the power of the  local pseudogaps depend on  $\alpha$, Eq. (\ref{power}),
    that critical value 
    $j_{c} (\alpha)$ depends on $\alpha$ as well. Accordingly, 
  magnetic moments remain unscreened even at $T=0K$, 
   when  $\alpha$ exceeds the
critical value 
\begin{equation} \label{alphafm}
\alpha_{\rm FM} = \alpha_{0 } + d j.
\end{equation}
 Out of the $N= L^d$
atomic sites in the system, magnetic moments remain free at all
temperatures if placed on one of $N_{FM}$
   sites, where a sufficiently strong local
pseudo gap is developed. Thus, for small $J$ there can be a macroscopic number of such
sites, although their density $n_{\rm FM} = N_{\rm FM}/N$, 
\begin{equation} \label{nfmt0}
n_{\rm FM} (T=0K) = L^{-d^2 \frac{
  j^2}{2  \eta }} n_{\rm M} ,
  \end{equation}
 vanishes,  $n_{\rm FM}  \rightarrow 0$ 
 for $L
\rightarrow \infty$.
  
  \subsection{Distribution Function of Kondo Temperature.}
  
     \subsubsection{   $T_{K} \rightarrow 0$ Limit of $P(T_{K})$} 
     \label{ptkto0}
  In deriving the conditional intensity, Eq. (\ref{cci}), and inserting it 
   in Eq. (\ref{tkc3}) we took 
    into account the correlations of all states  to the intensity 
   at the Fermi energy as characterised by   local pseudogaps and power law divergencies.
    Since their power  $r_{\alpha}$
     is distributed, we  can now obtain the distribution of  the Kondo temperature $T_{K}$ by 
solving Eq. (\ref{tkalpha}) for $\alpha( T_K )$ and
inserting it in $P(\alpha)$.
  To this end, we use  the Fourier representation of the delta-function.
   Then, we can  expand in  $F[\{ \alpha_{l} \}, T_{K}]$ and perform 
    the average, keeping $\alpha_{l} =\alpha$ at the  Fermi energy
     $E_{F}=E_{M}$ fixed.
     Next, we  use the conditional pair approximation 
     introduced above. 
     If we  impose the condition 
     $F[\{ \alpha_{l} \}, T_{K}]=1$ to obtain $T_{K}$ for given $\alpha$
      and insert this in $P(\alpha(T_{K}))$
      this  yields   in the limit of small $T_{K} \rightarrow \Delta$, 
        \begin{equation} \label{ptk0}
    P^0(T_{K}\rightarrow 0) \sim \left(  \frac{T_{K}}{E_{c}} \right)^{j-1}  L^{- \frac{
 (d j)^2}{2  \eta }}.
  \end{equation}
  For $j=0.25$ the 
   power of the $T_{K} \rightarrow 0$ tail is  $\beta = 1-j =0.75$  in exact agreement with the 
    numerical result in 3 dimensions reported in   Ref. \onlinecite{cornaglia}.
    However,  we find that its weight is vanishing with a power of the system size $L$.
      Note that the number of sites used in Ref. \onlinecite{cornaglia} is $N=2197$,
        yielding a level spacing $\Delta/T^{0}_{K} \approx 0.025$, so that Eq. (\ref{ptk0})
        indeed can explain the tail of the distribution  $T_{K} \le \Delta$ displayed in their Fig. 3.
     Clearly,  at larger $T_{K}$ the fluctuations of 
        the wave function intensities at energies away from the AMIT are important.
       These   can strongly change
         $T_{K}$ and the width of  its distribution as we  find in the next subsection.

       \subsubsection{  $ P(T_{K})$  for $T_{K}> \Delta$  }    
       
       In order to proceed in the calculation of $P(T_{K})$, 
     at $T_{K}$ exceeding the level spacing $\Delta$ we need to  take into account the 
        fluctuations of the intensities at  all energies. To this end,  
        let us  first rewrite Eq. (\ref{ptkexact})
         as
         \begin{eqnarray} \label{ptkft}
       P(T_{K}) =- \int_{0}^{\infty} d \alpha P(\alpha)   \langle   \frac{dF }{d T_{K}}
    \delta (1-F) \rangle_{\alpha},
       \end{eqnarray}
        where $\langle ... \rangle_{{\alpha}}$ denotes the average over all intensities,  
         except the one at the Fermi energy which is fixed to $L^{-\alpha}$. 
       Next,   we   use the Fourier representation of the 
  delta-function  to get
       \begin{eqnarray} \label{ptkft}
       P(T_{K}) =- \int_{0}^{\infty} d \alpha P(\alpha)   \frac{d}{d T_{K}} 
       \int_{-\infty}^{\infty} dt \frac{ i e^{it}}{2 \pi  t} 
       \nonumber \\ \times \langle \exp \left(  - i tF[ T_{K}]  \right) \rangle_{\alpha}.
       \end{eqnarray} 
              Now we can expand $F[\{ \alpha_{l} \}, T_{K}] $, 
                around $F[\alpha,T_{K}]=  \langle F \rangle_{\alpha} $
                to take into account the fluctuations of all intensities. 
   Expanding in  $ \delta F =
       F[\{ \alpha_{l} \}, T_{K}] -  F[\alpha,T_{K}]$, 
     and performing the integral over
        $t$, we find
               \begin{eqnarray} \label{ptkgeneral}
       & & P(T_{K}) =- \int_{0}^{\infty}  \frac{d \alpha  }{\sqrt{2 \pi } \Gamma } P(\alpha)  
     \frac{d F[\alpha,T_{K}] }{d T_{K}}   e^{ -\frac{ (1-F[\alpha,T_{K}])^2 }{ 2 \Gamma^2 } },
       \end{eqnarray}
       where $F[\alpha,T_{K}]=  \langle F \rangle_{\alpha} $, is given by
        Eq. (\ref{ftk})  and
       $\Gamma $ is defined by 
       \begin{equation}
      \Gamma^2  = \langle F^2 \rangle_{\alpha} - \langle F \rangle_{\alpha}^2,
       \end{equation}
        where $\langle ... \rangle_{\alpha}$ denotes the average over all $\alpha_{l}$,
         keeping only $\alpha$ at the Fermi energy fixed. 
           We 
         find,  that for $\Delta \le T_{K} \ll E_{c}$,
         \begin{eqnarray} \label{gamma}
         \Gamma^2  & & \approx  j^2 \frac{(T_{K}-\Delta)^2}{T_{K}^2} \left( c_{1} (\frac{T_{K} -\Delta }{E_{c} })^{-  \eta/d } 
          - c_{2}   \right. \nonumber \\  & & \left. +   5 \ln [\frac{T_{K} }{E_{c} }] -2  \ln [\frac{T_{K} }{E_{c} }]^2   \right),
          \end{eqnarray}
          where  we determined numerically, $c_{1} \approx 7.51$ and
          $c_{2} \approx 9.60$.
           Note that $\Gamma$ vanishes in the limit $T_{K} \rightarrow \Delta$,
           $\Gamma (T_{K} \rightarrow \Delta) =0$, 
            since then only energy levels in a range of order $\Delta$ around the Fermi energy 
             contribute, whose correlations are taken into account correctly by $F_{\alpha}$, already.
               Thus in this limit, the condition $1=F(\alpha,T_{K})$ is imposed
                exactly, and we recover the tail of the distribution,  Eq. (\ref{ptk0}), 
                 diverging with the power $\beta = 1-j$.
                 
             At larger $T_{K} > \Delta$, $\Gamma$ has a finite value and decays 
             for $\Delta  < T_{K} \ll E_{c}$ with the power $  \eta/d $.      
                $P(\alpha)$ is peaked at $\alpha = \alpha_{0}$ with a width that 
                scales with the system  size as $1/\sqrt{\ln L}$. Thus,  for $L \rightarrow \infty$,
                   $\alpha = \alpha_{0}$ is imposed 
                  in Eq.   (\ref{ptkgeneral}) for any finite $T_{K}$.
                  Thus we can substitute
 \begin{equation}
 \langle F[T_{K}] \rangle = j ( \frac{T_{K}-\Delta}{2 T_{K}} + \ln \frac{D}{2 {\rm Max} (T_{K},\Delta)}),
 \end{equation}
  and find that    the  distribution diverges at  $\Delta < T_{K} \ll T_K^{(0)} $ as
        \begin{eqnarray} \label{ptktail}
        P(T_{K})&& \approx \left( \frac{T_{K}}{E_{c}}\right)^{ \frac{ \eta}{2 d } -1}
        \times
        \nonumber \\ &&
         \exp \left\{- \frac{1}{2 c_{1}} \left( \frac{T_{K}}{E_{c}}\right)^{\eta/d} \ln^2 \left[ \frac{T_K}{T_K^{(0)}} \right]
  \right\}.
        \end{eqnarray}
       with the power  $\beta =1-  \eta/d /2 $. 
         In $d=3$ dimensions,  with $\alpha_{0}=4$, the power is $\beta_{d=3}=2/3$ which is smaller
         than the one obtained numerically in Ref. \onlinecite{cornaglia},
          $\beta \approx 0.75$.   We  note, that           
          there is a noticeable   
           deviation towards smaller powers for $T_{K}> \Delta \approx 0.03$ in 
            the Fig. 3 of Ref. \onlinecite{cornaglia}.

\section{Kondo Effect in the Metal Phase }          
\label{sec:metallic}

\begin{figure}[h]
\includegraphics[width=9cm,angle=0]{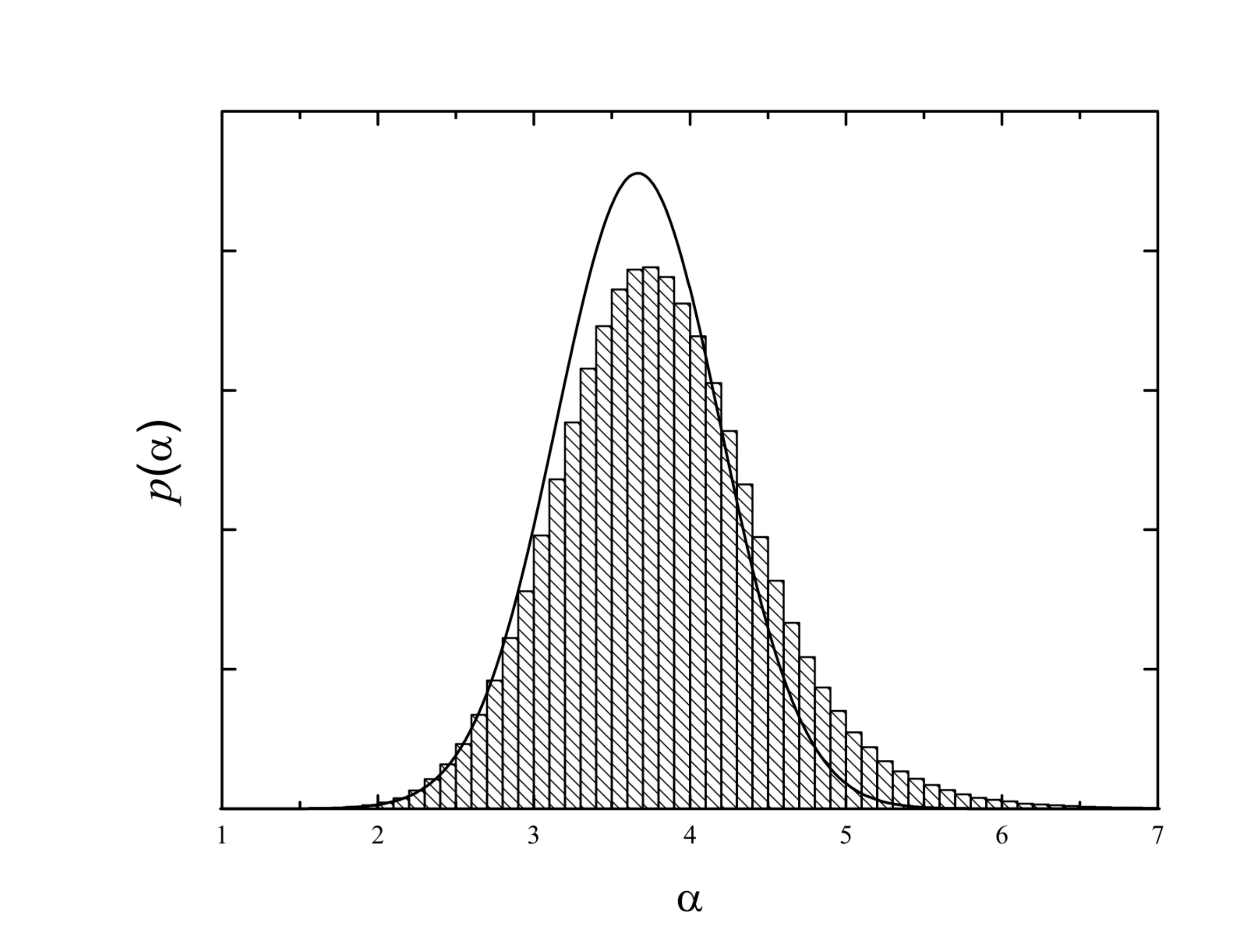}
\caption{The distribution of $\alpha$ defined in the metallic regime by Eq. (\ref{alphametal}), as
obtained by exact numerical diagonalisation of a 3D sample  of size 
 $L^3= 128^3 $ in units of  the grid cell volume
   $a^3$.
 The energy of that state is approximately  $E=0$,  and a box distribution of uncorrelated disorder
  potential with  $W=15 t$ is taken. The analytical expression 
   Eq. (\ref{palphametallic}) is plotted for comparison as the solid line for a correlation  
    length $\xi = 22 a$ 
    (taken from the numerical results of Ref. \onlinecite{kmkprl}).  }
\label{fig:palphametal}
\end{figure}

\begin{figure}[h]
\includegraphics[width=9cm,angle=0]{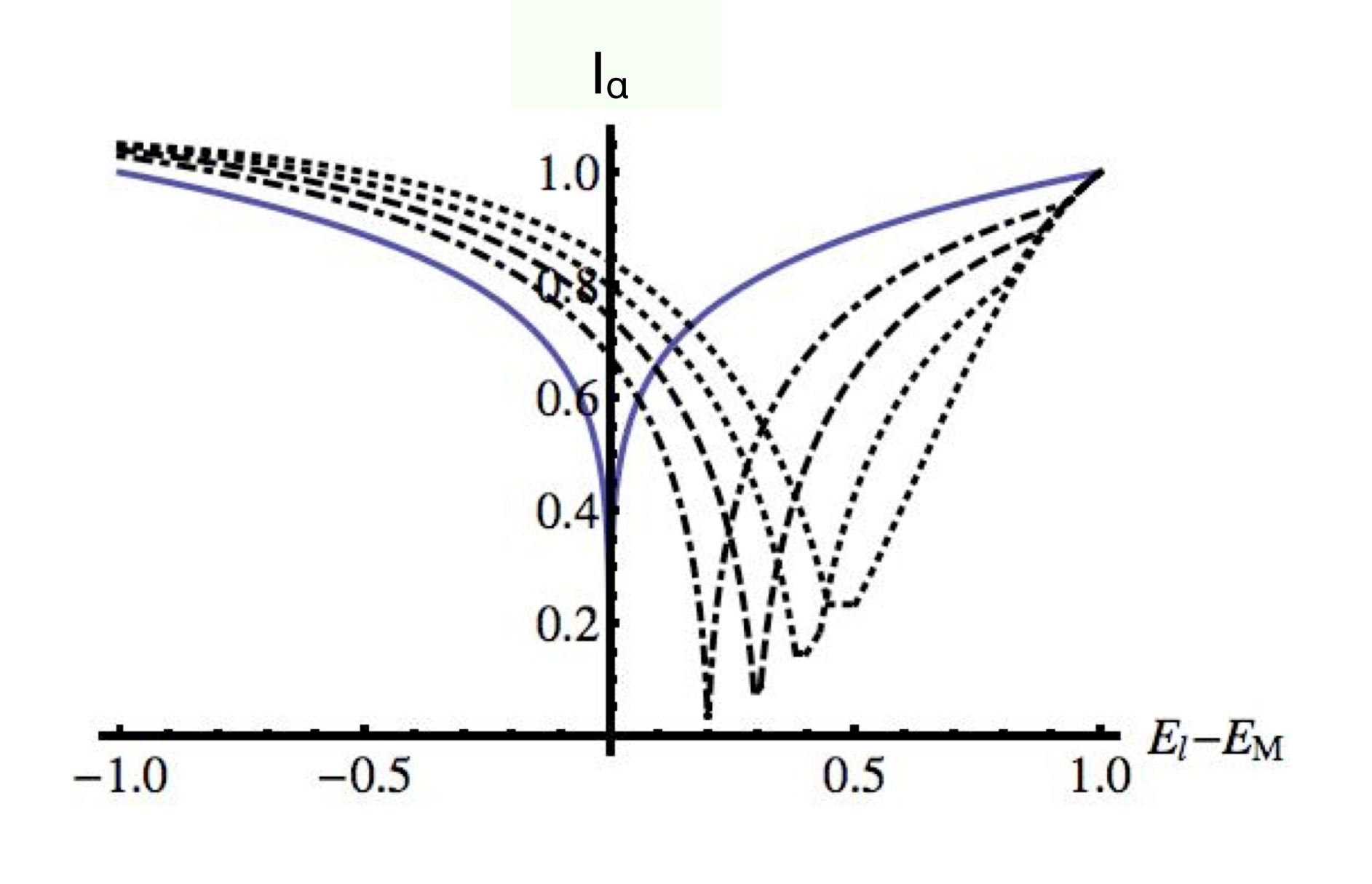}
\caption{(Color online) The conditional intensity $I_{\alpha}$ at energy $E_{l}$,
given that
a state at the energy $E_{k}$ has intensity $L^d |\psi_{k} ({\bf
  r})|^2 = \xi_{k}^{d-\alpha}$,
 relative to the intensity of an
  extended state. It is plotted as function of the distance of energy $E_{l}$ from
  the mobility edge energy $E_{M}$, in units of the correlation energy
  $E_{c}$. The exponent $\alpha$  at the energy $E_{k}$, takes the value $\alpha = 4.5$. The 
   energy $E_{k}$ is varied from the mobility edge into the metallic regime
  with $(E_{k} -E_{M})/E_{c}= 0, 0.2, 0.3, 0.4, 0.7$, from left to
  right. Instead of a local pseudo gap, one sees an increasingly
  shallow depression. Here we set $d=3$ and
  $\alpha_{0}=4$.  Note that it varies slowly in a range $\Delta_{\xi_{k}}$
   around the Fermi energy. }
\label{fig:pseudogapmetal}
\end{figure}

\begin{figure}[h]
\hspace{-0cm}
\includegraphics[width=9cm,angle=0]{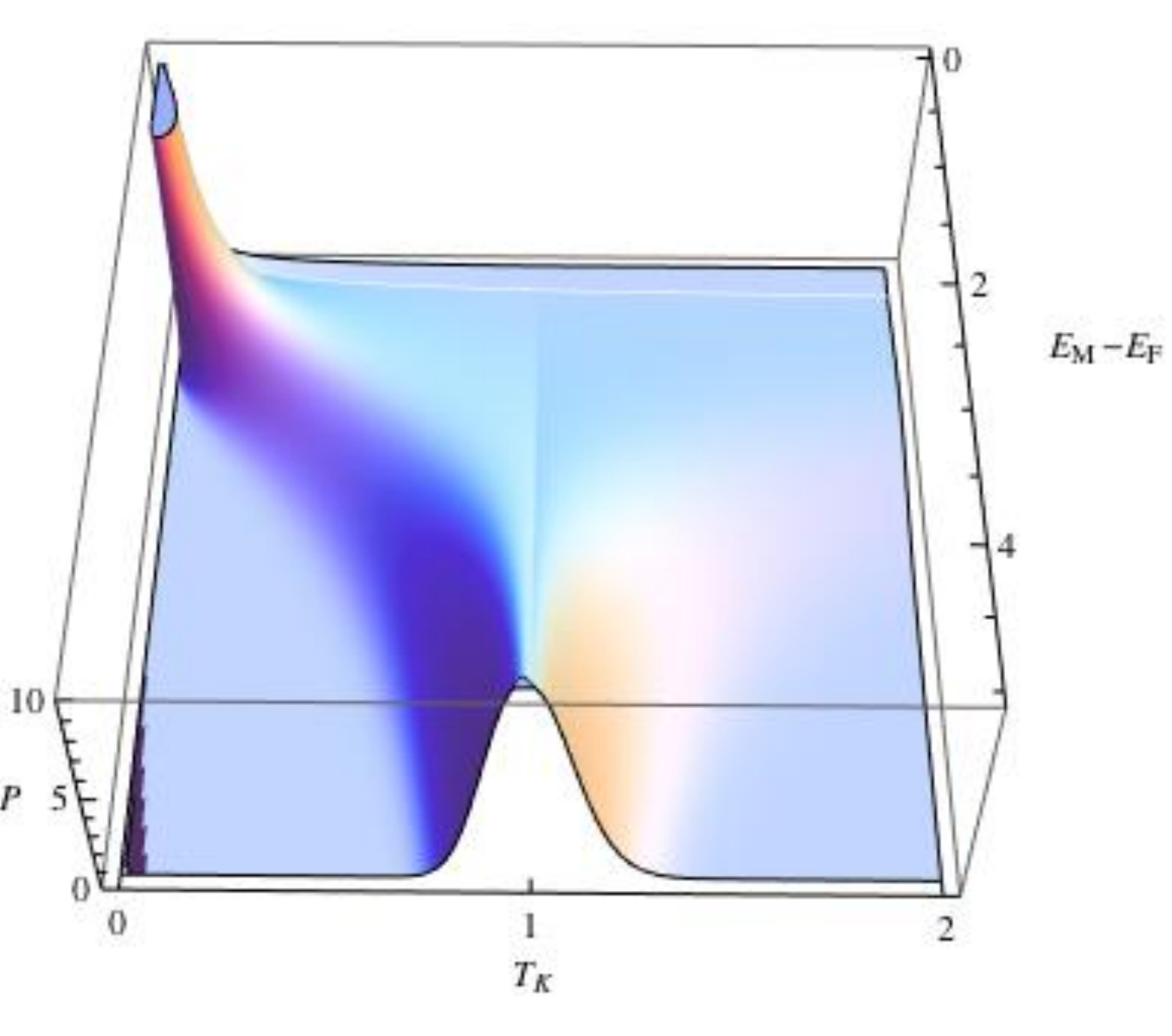}
\caption{(Color online) The distribution of Kondo temperatures $T_{K}$ in units of
  $T_{K}^{(0)}$ in the metallic phase, 
   Eq. (\ref{ptkm}), is plotted as function of the distance to the mobility edge,
  $E_{F}-E_{M }$ in units of $E_{c}$ for an exchange coupling $j= 1/5$. }
\label{fig:ptkmetal}
\end{figure}

\subsection{Multifractality in the Metallic Phase}
In the metallic regime all wave functions are extended and their
intensities scale with the inverse system volume,  $|\psi|^2 \sim L^{-d}$.
 On length scales
smaller than the correlation length $\xi$, multifractal fluctuations
of the wave function intensity  occur as long as $\xi$ is larger
than the microscopic length scale $a_{c}$.\cite{cuevas,ioffe2}
 As pointed out in Refs. \onlinecite{ioffe2,mirlin} the  moments of the intensity
  do scale  with the 
  correlation length $\xi$ as 
  $L^{d q} \langle | \psi|^{2 q} \rangle \sim \xi^{(d-d_{q})(q-1)}$.
Therefore, in the metallic
regime  we define  $\alpha$ as 
\begin{equation} \label{alphametal}
L^d |\psi_{l} ({\bf r})|^2 =
\xi_{l}^{d-\alpha_{l}},
\end{equation}
 where $\xi_{l}$ is the correlation length of
state $l$.  Notice that this definition of $\alpha$ crosses over to the one we
used above in the critical regime, where $\xi$ diverges and is
replaced by the system size $L$,  when  $L < \xi$. It has  to a good approximation
  still the  Gaussian distribution, 
\begin{equation} \label{palphametallic}
P(\alpha_{l}) \sim \exp [ - \ln
  \xi_{l} \frac{(\alpha_{l}-\alpha_{0})^2}{2   \eta }],
  \end{equation}
   where its
width scales with the logarithm of the correlation length $\xi_{l}$.
  This is confirmed in Fig. \ref{fig:palphametal}, where we plot Eq. (\ref{palphametallic})
  together with the numerical result, as obtained from exact diagonalisation. 

\subsection{Intensity Correlations in the Metal}
There are still power-law correlations
in energy between wave function intensities as given by Eq. 
 (\ref{criticalcorrelations2}).   Averaging with the conditional
distribution function Eq. (\ref{eq:Ptwom}) in the metallic regime, we find the
conditional intensity in the metallic regime:
 given that
a state at the energy $E_{k}$ has intensity $L^d |\psi_{k} ({\bf
  r})|^2 = \xi_{k}^{d-\alpha}$, a state at energy $E_{l}$ at position
${\bf r}$ has on average the intensity
\begin{equation}
\label{intensitymetallic}
I_{\alpha {\rm }} (\xi_{l},\xi_{k})= \langle L^d |\psi_{l}
({\bf r})|^2 \rangle_{\alpha} = K_{lk}^{\frac{\alpha-\alpha_{0}}{d}
  -\frac{  \eta }{2d^2} \frac{ \ln K_{l k} }{ \ln \xi_{k}} },
\end{equation}
where $K_{{lk}} = {\rm Max} \left\{ | E_{l} -E_{k} |, {\rm Min} (
\Delta_{\xi_{l}}, \Delta_{\xi_{k}}) \right\}/E_{c} $ and
\begin{equation}
\Delta_{\xi_{l}} = E_{c} (a_{c}/\xi_{l})^d,
\end{equation}
 is the mean level spacing
of a sample of finite size $\xi_{l}$ (for the derivation, see Appendix
\ref{sec:A}). Thus, the intensity at the energy $E_{l}$ has still a
dip when $\alpha > \alpha_{0}$, although the LDOS at energy $E_{l}
\rightarrow E_{k}$ is no longer suppressed to zero but rather to a
finite value (see Fig. \ref{fig:pseudogapmetal}), given by, 
\begin{equation} \label{ldosmetallic}
\langle L^d |\psi_{l} ({\bf r})|^2 \rangle_{\alpha}|_{E_{l}
  \rightarrow E_{k}} = \xi_{k}^{
  (d- \alpha)} = \left|\frac{E_{k}-E_{M}}{E_{c}} \right|^{\nu
  (\alpha-d)},
\end{equation}
 which is  slowly varying with energy in an interval of order $\Delta_{\xi_{k}}$ 
 around the Fermi energy. 
 
\subsection{Distribution of Kondo Temperatures }
\label{sec:distrbKondometal}

Since the intensity is finite at all sites in
the metallic regime we expect that the 
  magnetic impurity spin is at low temperatures always screened. Therefore,  the 
 Kondo temperature does not vanish in the metallic regime at any site. 
   According to Eq. (\ref{ldosmetallic}) the local intensity can be substantially 
   suppressed or enhanced, depending on the random value of $\alpha$.
   Its distribution, as given by Eq. (\ref{palphametallic}), has a finite width.
     On the other hand, the second moment of $F$ does saturate to a finite value 
      when $T_{K} < \Delta_{\xi}$:
            \begin{equation} \label{gammametal}
         \Gamma^2|_{T_{K} < \Delta_{\xi}} 
        \rightarrow  j^2 c_{1} (\frac{\Delta_{\xi} }{E_{c} })^{-  \eta/d } ,
          \end{equation}
          where   $c_{1} \approx 7.51$.
   Inserting both Eqs. (\ref{gammametal},\ref{palphametallic})
    into the expression for $P(T_{K})$, Eq. (\ref{ptkgeneral}) we find
       that  the probability to find a 
        small Kondo temperature such that  $T_{K} < \Delta_{\xi} = |(E_{F}-E_{M})/E_{c}|^{\nu d}$,
       is decaying to zero as
       \begin{eqnarray} \label{ptktailmetal}
        P(T_{K})&& \approx \left( \frac{\Delta_{\xi}}{E_{c}}\right)^{ \frac{ \eta}{2 d }} \frac{1}{T_{K}}
         e^{ - \frac{1}{2 c_{1}} \left( \frac{\Delta_{\xi}}{E_{c}}\right)^{\eta/d} \ln^2 \left( \frac{T_K}{T_K^{(0)}} \right)
  }.
        \end{eqnarray}
        Thus, in the metallic phase
       $E_{F}> E_{M}$ there are no free magnetic moments and the low $T_{K}$-tail 
         terminates at $T_{K} \approx \Delta_{\xi}$ as seen in Fig. (\ref{fig:ptkmetal}). 
       For $T_{K} > \Delta_{\xi} $ a power law tail of the distribution can still be observed,
       as seen in Fig.  (\ref{fig:ptkmetal}).
            There, we plot  $P(T_{K})$
             as obtained by Eq. (\ref{ptktail})  and substituting 
              $T_{K}$ by ${\rm Max} (T_{K},\Delta_{\xi})$ in the expression for 
               $\Gamma$, Eq. (\ref{gamma}),
              \begin{eqnarray} \label{ptkm}
        P(T_{K})&& \approx \left( \frac{Max(T_{K},\Delta_{\xi})}{E_{c}}\right)^{ \frac{ \eta}{2 d } }
         \frac{1}{T_{K}}
        \times
        \nonumber \\ &&
         \exp \left\{- \frac{1}{2 c_{1}} \left( \frac{Max(T_{K},\Delta_{\xi})}{E_{c}}\right)^{\eta/d} \ln^2 \left[ \frac{T_K}{T_K^{(0)}} \right]
  \right\}.
  \end{eqnarray}

 %

\section{Kondo Effect in the Insulator }
\label{sec:insulator}

In the insulating regime each localized state is restricted to a
volume of the order of $\xi_{l}^d = \xi_{c}^d (E_{l})$,  the
localization volume. On length scales smaller than the localization
length $\xi_{l}$ there are still multifractal fluctuations. In
addition, the wave functions in the insulating regime are 
power-law correlated close to the AMIT.\cite{cuevas} Neglecting a
small logarithmic enhancement which occurs at energy spacings smaller
than the local level spacing $\Delta_{\xi} = \xi^{-d} D$,\cite{cuevas} we can get
the distribution of $T_{K}$  simply by using the results
obtained at the critical point and replacing the system size $L$ by
the localization length $\xi_{c}$. Defining 
\begin{equation}
|\psi_{l}(x)|^2 =
\xi_{l}^{-\alpha},
\end{equation}
 we thus find that the distribution function of
$\alpha$ within a localization volume is the same as if we had
considered the distribution at the AMIT in a finite volume of order
$\xi_{l}^d$. We note that the probability to find a state at energy
$E_{l}$ inside the localization volume is decaying with the system volume 
 $L^d$ as $\xi_{l}^d/L^d$.
Outside the localization  volume the intensity decays exponentially,
$|\psi_{l}({\bf r})|^2 \sim \exp[-2 r/\xi]$ corresponding to $\alpha
= 2 r/(\xi_{l}\ln \xi_{l}) $. 
 Since  most sites in the sample are a distance  $r \sim L \rightarrow
\infty$ away from the localization volume one finds $\alpha \sim L \rightarrow \infty$
 at most sites of the sample.  
%
The correlation function between two wave functions at different
energies $E_{l}$ and $E_{k}$ is  still given by
Eq. (\ref{criticalcorrelations}) where $\Delta_{\xi_{l}}$ is now the
local level spacing at energy $E_{l}$. Accordingly, the joint
distribution function has the form given in
Eq. (\ref{eq:Ptwom}). However, the difference to the metallic regime
is that there are only discrete energy levels which are separated by  the local level
spacing $\Delta_{\xi_{l}}$. This difference is important, especially
when calculating the Kondo temperature, since the hard gap
$\Delta_{\xi_{l}}$ in the insulating regime cuts off the Kondo renormalization flow at small
energies.\cite{meraikh} 

   Thus, we can conclude that 
    there is a finite density of free magnetic moments  $n_{FM}$, which remain unscreened. 
 Therefore,  we need to subtract this density from the distribution function
of $T_{K}$  given by Eq. (\ref{ptkgeneral}).  The 
 width  $\Gamma$ is given by Eq. (\ref{gamma}) when
  substituting $\Delta$ by the level spacing in a localization volume, $\Delta_{\xi}$
   for $T_{K} \ge \Delta_{\xi}$,
    \begin{eqnarray}
         \Gamma^2  & & \approx  j^2 \frac{(T_{K}-\Delta_{\xi})^2}{T_{K}^2}  c_{1} (\frac{T_{K} -\Delta_{\xi} }{E_{c} })^{-  \eta/d },
          \end{eqnarray}
          where  $c_{1} \approx 7.51$.
     For $T_{K} < \Delta_{\xi}$, we find that it vanishes, $\Gamma =0$, 
      and the condition $1-F_{\alpha}$ is
      enforced exactly. Therefore, we get from Eq. (\ref{ptkgeneral}) the low $T_{K}$ tail in the localised regime as
\begin{equation}
\label{ptkxi}
P(T_K) = \left(1- \frac{n_{FM} (T=0)}{n_{M}}\right)
  \left( \frac{E_c}{T_K} \right)^{1-j} \xi^{- \frac{1}{2  \eta }
  (d j)^2}.
\end{equation}
Thus, we find that 
 the distribution diverges 
 with the  power $ \beta =1-j$.
   This is in full agreement with the numerical results of Ref. \onlinecite{cornaglia},
    where a power $0.75$ has  been obtained for $j=0.25$ for a wide range of disorder amplitude 
     $W$. 
    Also, the  increase of the weight of the power law tail  with 
      disorder strength $W$ is in good agreement with the numerical results. 


The finite density of free magnetic
moments is found to be,\cite{kettemann}
\begin{eqnarray}
\label{nfm}
n_{\rm FM}(T=0K) &=& n_{M} \xi^{-\frac{
    1}{2   \eta } (d j)^2} \nonumber \\ &=&  n_{M}  \left( \frac{ W -W_{c}}{W_{c}} \right)^{\frac{\nu
    }{2   \eta } (d j)^2},
\end{eqnarray}
which decays to zero as a power law when the disorder amplitude $W$ 
approaches the AMIT at $W_{c}$. It converges to the total density of
magnetic moments $n_{M}$  far away from the mobility edge, $E_{M}
-E_{F} \rightarrow E_{c}$. Thus, according to this expression all
magnetic moments should be free in the strongly localized regime 
where $\xi \rightarrow a_{c}$.

\section{  Kondo Effect 
 in 2D Anderson Insulators} 
 \label{sec:insulator2D}
 We can apply this analysis also to
two-dimensional disordered electron systems, where all states are
localized. The Kondo effect in such systems has been studied
numerically based on the 1-loop equation,\cite{kettemannjetp,cornaglia} 
 and with nonperturbative methods in
Ref. \onlinecite{zhuravlev}, where both the distribution of Kondo
temperatures and the density of free magnetic moments have been
obtained. The two-dimensional localization length in the absence of a
magnetic field is known to depend exponentially on the disorder
strength, $\xi_{2D} = g \exp (\pi g)$, where $g = E_F \tau$. The
scattering rate $1/\tau$ is related to the disorder amplitude $W$ as
$1/\tau = \pi W^2/6 D$. There are weak wave function correlations in two
dimensions which  are logarithmic, with an  amplitude of order $1/g$. For  weak disorder, 
 $g \gg 1$ we
can rewrite this correlation as an effective power law with power
\begin{equation}
  \eta_{2D} = 2/ \pi g.
\end{equation}
 The correlation energy in 2D  is of the order of
the elastic scattering rate, $E_{c~2D} \sim 1/\tau$.  Thus, for
systems whose size $L$ is smaller than the localization length
$\xi_{{2D}}$, the 2D system behaves like a critical
system with $\alpha_{0}$ defined by 
  $  \eta _{2D} = 2/ \pi g = \alpha_{0}-2 $, 
   or 
   \begin{equation}
   \alpha_{0} = 2 + \frac{2}{\pi g}. 
   \end{equation}
  There is a critical exchange coupling $J_c^{(1)}$
above which there is no more than one free magnetic moment in the
whole sample.\cite{kettemann} Substituting $  \eta_{2D}$, we find, that 
\begin{equation}
J_c^{(1)} = \sqrt{\frac{D}{3 E_F }} W.
\end{equation}
 This is in good agreement
with 
Ref. \onlinecite{zhuravlev} where $J_c^{(1)}$ has been 
 determined numerically for a 2D disordered system, and found 
  to increase linearly with disorder amplitude $W$. There, 
   and in Ref. \onlinecite{cornaglia} the density of free moments 
    have been obtained which we can now compare with our 
     analytical expression, 
     \begin{equation} 
     n_{\rm FM}(T=0K) = n_{M} \xi_{2D}^{-\frac{
    d}{  \eta_{2D}} j^2} = n_{M} (g \exp  (\pi g) )^{ -\pi g j^2},
     \end{equation}
      with $ g = 6/(\pi W^2)  E_{F}/D$.
     By
substituting $\xi_{2D}$ into Eq. (\ref{ptkxi}) and inserting the
parameters used in Ref. \onlinecite{zhuravlev} we compared the
analytical distribution of $T_K$ with the numerical results and found
a good qualitative agreement (not shown). We note that 
 there is a low $T_{K}$ power-law tail  divergence  where the power is given by $\beta = 1-j$.

\section{Zero Temperature Quantum Phase Diagram}
\label{sec:quantumphase}

It is well known that the scattering of conduction electrons by
magnetic impurities can lead to the relaxation of the conduction
electron spin, and thereby the loss of the electron phase
coherence. At finite temperature this leads to a suppression of the
quantum corrections to the conductance, the so-called weak
localization corrections. In the low-temperature limit phase
coherence is restored, but the magnetic scattering may still break the
time-reversal symmetry of the conduction electrons similarly to an
external magnetic field. The breaking of time-reversal symmetry is
known to weaken Anderson localization and thereby the localization length
becomes enhanced. In systems with an Anderson metal-insulator
transition the transition is shifted toward stronger disorder
amplitudes $W$ and lower electron density
$n$.\cite{larkin,wegner1986,droese} Thus, the symmetry class 
changes, shifting the AMIT from the {\it orthogonal} symmetry class
(time-reversal symmetric) to the {\it unitary} symmetry class (broken
time-reversal symmetry).\cite{larkin,wegner1986,droese}

In the presence of an external magnetic field this change of the
symmetry class of the conduction electrons from orthogonal to unitary
is governed  by the parameter $X_{B} =\xi^2/l_{B}^2$, where
$l_{B}$ is the magnetic length. Therefore, the spin
scattering rate due to magnetic impurities $1/\tau_s$ is expected to
enter through the symmetry parameter $X_s = \xi^2/{ D_{e}} \tau_s $,
where ${D_{e}}$ is the diffusion constant and $\xi$ is the correlation
(localization) length on the metallic (insulating) side of the
AMIT.\cite{hikami} When $X_{s} \ge 1$, the electron spin relaxes
before it can cover the area limited by $\xi$ and the system is in the
{\it unitary regime}. One can then study the crossover of the mobility
edge through a scaling Ansatz for the conductivity on the metallic
side, as done in Ref.  \onlinecite{larkin} in the case of a magnetic
field. Following this approach, using the spin scattering rate
$1/\tau_{s}$, we get
\begin{equation}
\sigma (1/\tau_{s}) = \frac{e^2}{h \xi} f (X_{s}).
\end{equation}
The conductivity then goes to zero at the critical disorder $W^M_{c}
(1/\tau_{s})$ as $\sigma (1/\tau_{s}) \sim (W^M_{c}
(1/\tau_{s})-W)^{\nu}$, where the index $M$ indicates that this is the
critical disorder strength when it is approached from the metallic
side, $W < W^M_{c} (1/\tau_{s})$.
      
Coming from insulating side, we need instead to apply the scaling
Ansatz to the dielectric susceptibility $\chi_{e}$,\cite{dielectric,leeramakrishnan}
\begin{equation} \label{chi}
\chi_{e} (1/\tau_{s}) = \xi^2 g (X_{s}).
\end{equation}
This diverges at the critical disorder $W^I_{c} (1/\tau_{s})$ as $\chi
(1/\tau_{s}) \sim (W- W^I_{c} (1/\tau_{s}))^{-2 \nu}$, where the index
$I$ indicates that this is the critical disorder strength when it is
approached from the insulating side, $W >W^I_{c} (1/\tau_{s})$. While
in a magnetic field these two critical points are found to
coincide,\cite{droese} we will see below that, in the presence of
magnetic impurities, $W^M_{c} (1/\tau_{s})$ and $W^I_{c}(1/\tau_{s})$
can be different due to the Kondo effect.

When a finite concentration of classical magnetic impurities $n_{M}$
with spin $S$ is present, the magnetic relaxation rate at zero
temperature is given by $1/\tau^{\rm classical}_{s} = 2\pi n_{M} S^2
j^2 \rho (\epsilon_F)$, where $\rho (\epsilon_F)$ is the density
of states at the Fermi energy. However, the quantum mechanical nature
of the impurity spins affects this rate in several ways: First, its
magnitude is enhanced since the quantum mechanical eigenvalue of the
square of the spin is $S(S+1)$. This  results for $S=1/2$ in a factor of
$3$ enhancement. Secondly, the Kondo effect tends to screen the
impurity spin leading to a vanishing spin relaxation rate at zero
temperature when magnetic impurities are dilute. However, at finite
temperature the Kondo correlation can instead enhance the spin
relaxation rate with a maximum at $T_K$. This effect has been
observed in weak-localization experiments as a plateau in the   temperature dependence of the 
dephasing
time.\cite{bergmann,mw00} Recently, its full
temperature dependence was obtained
numerically.\cite{zarand,micklitz} A good agreement with the
numerical results can be obtained through the approximate expression,
\cite{kettemannjetp}
\begin{eqnarray}
\label{eq:tauapprox}
\frac{1}{\tau_s^{(0)}} (T) & = & \frac{\pi\, n_m\, S(S+1)}{\rho} \left\{
\ln^2 \left( \frac{T}{T_K} \right) \right. \nonumber \\ & & \left. +
\pi^2 S (S+1) \left[ \left( \frac{T_K}{T} \right)^2 + \frac{1}{\beta}
  - 1 \right] \right\}^{-1},
\end{eqnarray}
with $\beta = 0.2$, as obtained numerically.\cite{zarand} The
temperature dependence scales with $T_K$. Note that in the
low-temperature limit the spin relaxation rate vanishes as
$T^2/T_{K}^2$, similarly to the  inelastic scattering rate in a
Fermi liquid and in agreement with Nozieres' renormalized Fermi
liquid theory of dilute Kondo systems. \cite{nozieresfl}

\subsection{ Metal Phase} In the zero-temperature limit, coming
from the metallic side of the AMIT, the Kondo screening results in the
vanishing of the spin relaxation rate, $1/\tau_{s}(T=0) = 0$.\cite{metalspinrelaxation}
Therefore, $X_{s}=0$ and the AMIT occurs at the orthogonal critical
value for time-reversal symmetric systems, $W_{\rm c}^{\rm O}$. For
the Anderson tight-binding model in three dimensions, one finds
$W_{\rm c}^{\rm O}/t = 16.5 \pm 0.02 $, with the orthogonal critical
exponent $\nu_{\rm O} = 1.57 \pm 0.02$ and the correlation exponent
$  \eta/d _{\rm O} = 0.56 \pm
0.02$.\cite{ohtsukislevinreview,rmpmirlinevers}. Most recently with a more  accurate
multifractal scaling method  $W_{\rm c}^{\rm O}/t =  16.530(16.524, 16.536)$,
 $\nu_{\rm O} =1.590(1.579, 1.602)$, $\alpha_{0} = 4.048(4.045, 4.050)$ and
$  \eta _{\rm O} = 1.763(1.792, 1.727)$ was obtained\cite{roemer}.

Looking now at the transition point we can conclude from the
analytical results of Sec. \ref{sec:insulator}, namely
Eq. (\ref{nfm}), that at the AMIT the density of free magnetic moments
vanishes. However, there can remain a macroscopic number of free
moments. We find that from the $N_{M}$ magnetic moments in the sample,
at least $\sqrt{N_{M}}$ of them remain free when the exchange coupling
does not exceed the value $j_c^{(2)} = \sqrt{  \eta/d } $.

\begin{figure}[h]
\includegraphics[width=8.2cm]{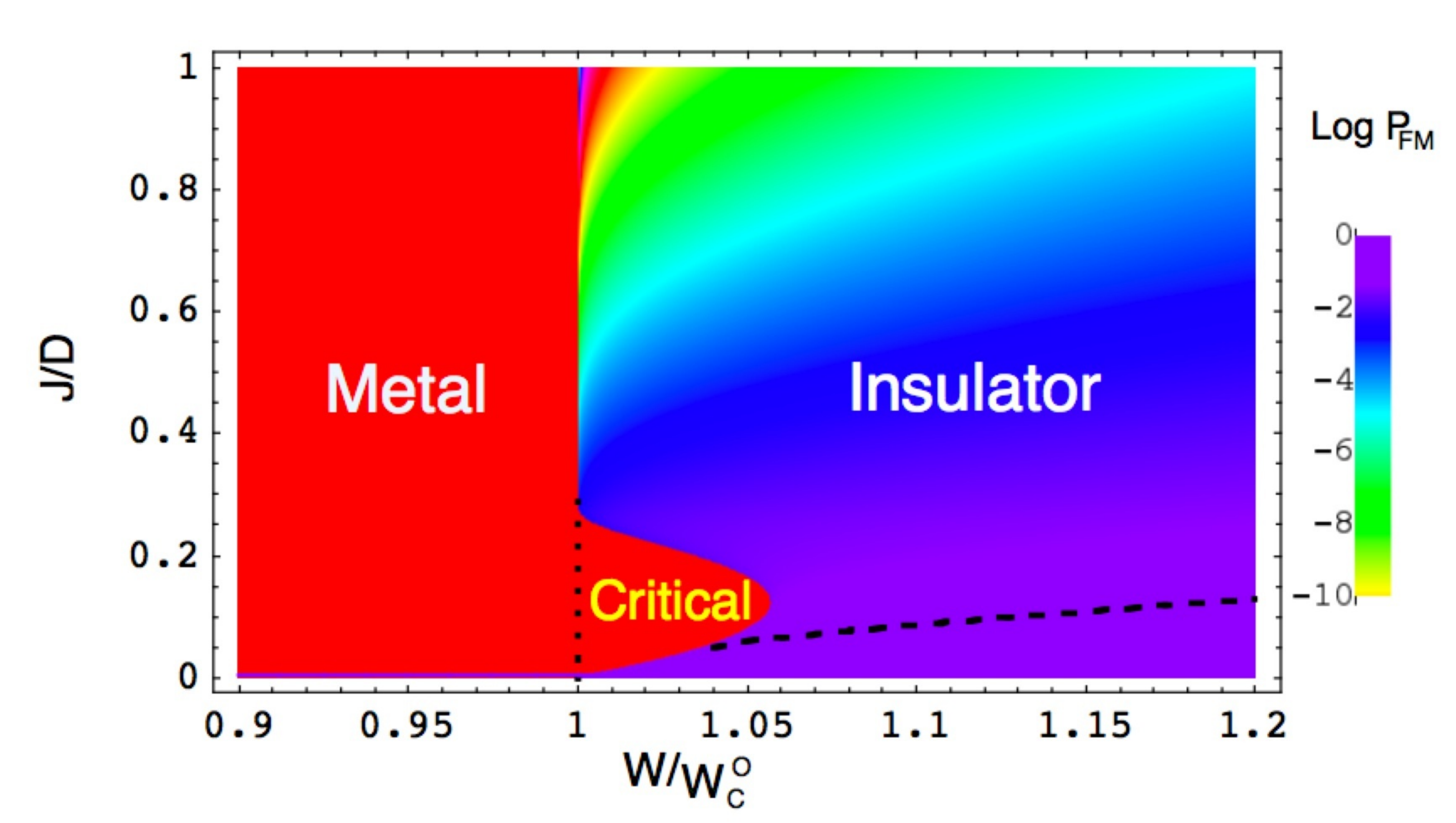}
\caption{(Color online) The fraction of free magnetic moments $P_{\rm
    FM}$ at $T=0K$, Eq. (\ref{nfm}), in a three-dimensional disordered metal as
  function of the exchange coupling $J$ (in units of the band width
  $D$) and disorder strength $W$ (in units of the critical value
  $W_{\rm c}^{\rm O} $). Critical correlations result in a finite
  $P_{\rm FM}$ even for large $J > J_c^A$ (dashed line]). For $J <
J^\ast $, Eq. (\ref{jstar}) there is a critical region for disorder
amplitudes $W_{\rm c}^{\rm O} < W < W_{\rm c} (J)$, where $W_c(j=J/D)$ is
given by Eq. (\ref{wcj}). This figure was previously published by
the authors in  Ref. \onlinecite{kettemann}.}
\label{fig:phasediagram}
\end{figure}
 
\subsection{ Insulator Phase} On the insulating side of the AMIT the
density of free magnetic moments is finite as given by
Eq. (\ref{nfm}).  Therefore, the time reversal symmetry and the spin
symmetry of the conduction electrons is broken in proportion to the
spin relaxation rate $1/\tau_{s}$. This leads to a shift of the AMIT
from $W_{c}^{\rm O}$ to $W^I_{c} (1/\tau_{s})$, which we will
determine in the following.

Before proceeding we note that in a disordered system the spin
relaxation rate is proportional to the local density of states
$\rho(E,{\bf r})$ at position ${\bf r}$, since it is given by
\begin{eqnarray}
\label{tausdisordered}
\frac{1}{\tau_s (E,T,{\bf r})} & = & - \, n_m\, \Omega
\frac{\rho(E,{\bf r})}{\nu} \left( t_0^2\, {\rm Im}\, G_d + \pi \rho
\, t_0^4\, G_d^2 \right) \nonumber \\ & = & \frac{\rho(E,{\bf
    r})}{\nu} \frac{1}{\tau_s^{(0)} (E,T)},
\end{eqnarray}
with the propagator of a localized d-level in the Anderson model
$G_d$.\cite{zarand} Here, $1/\tau_s^{(0)}$ is the spin scattering rate
in a clean system as given by Eq. (\ref{eq:tauapprox}). Thus,
Eq. (\ref{tausdisordered}) leads us to conclude that the spin
scattering rate not only depends on the ratio $T/T_K$ but also
explicitly depends on the LDOS. Since both $T_K$ and $\rho(E,{\bf r})$
are randomly distributed, $1/\tau_s(E, T, {\bf r})$ is distributed as
well.

In Section \ref{sec:insulator} it was found that on the insulating side of the AMIT
the LDOS at the position of unscreened magnetic moments scales as
$\rho (\epsilon_F, {\bf r}) \sim \xi^{d-\alpha(J)}$, where $\alpha_{FM}
= \alpha_0+ d j$ is the critical value.
 Therefore,  for $\alpha > \alpha_{FM}$ the magnetic moments remain
free. According to Eq. (\ref{tausdisordered}) the spin
relaxation rate due to the free moments depends itself on the
localization length as
\begin{eqnarray}
  \frac{1}{\tau^{\rm FM}_s}(T=0) & = & 2\pi\, n_{\rm FM}(\xi)\, S
  (S+1) j^2 \times \nonumber \\ & &  \rho
  (\epsilon_F) \left(\frac{\xi}{a_{c}}\right)^{d-\alpha_{0}-d
    j},
\end{eqnarray}
where the density of free moments is given by Eq. (\ref{nfm}) and,
 depends on the localization length $\xi$ as $n_{\rm FM} =
n_{M} (\xi/a_{c})^{-\frac{ d}{2   \eta/d } j^2}$.
Following Ref. \onlinecite{larkin}, we get the critical disorder
amplitude when the argument of the scaling function $g$ in
Eq. (\ref{chi}) is of order unity, yielding the condition $D \tau_s(J)
= \xi [W_c^I(J) -W_{\rm c}^{\rm O}]^2$ as function of the exchange
coupling $J$. This condition is valid as long as the deviation from $W_{\rm
  c}^{\rm O}$ is small. For larger deviations it will converge to its
unitary value $W_{\rm c}^{\rm U}$. The exact analytical form
cannot be obtained from this phenomenological scaling approach. We
note that, in contrast to the case of an external magnetic field, as
considered in Ref. \onlinecite{larkin}, according to
Eq. (\ref{taufm}), $\tau_{s}$ depends on the localization length $\xi$
as
\begin{equation}
\label{taufm}
\frac{1}{\tau^{\rm FM}_s}(T=0) = \frac{1}{\tau^{0}_s}
\left(\frac{\xi}{a_{c}}\right)^{- \frac{d}{2   \eta/d }(
  j+  \eta/d )^2 } ,
\end{equation}
where 
\begin{equation} \label{ts0}
\frac{1}{\tau^{0}_{s}} = 2\pi n_{M} S (S+1) j^2 \rho (\epsilon_F)
\end{equation}
Thus, we finally get the shift of the
critical disorder as function of $J$ as
\begin{equation}
\label{wcj}
W^I_{\rm c} (j)= W_{\rm c}^{\rm O} + W_{\rm c}^{\rm O} \left(
\frac{a_{c}^2}{D_{e} \tau^{0}_{s}}\right)^{\kappa(j)} ,
\end{equation}
where $1/\kappa(j) = \nu [2-\frac{d^2}{2   \eta }( j +  \eta/d )^2]$.
This result is valid for small deviations from $W_{\rm c}^{\rm O}$.
For larger deviations it will approach the unitary value $W_{\rm
  c}^{\rm U}$ in a still unknown form. We see that for $j = j^\ast$
  where 
  \begin{equation} \label{jstar}
  j^\ast =
 \frac{2 \sqrt{  \eta}}{d} -  \frac{\eta}{d},
\end{equation}
 which gives in $d=3$,    $j^\ast \approx 0.276$,
 the exponent $\kappa(j)$
diverges and $W^I_{\rm c}$ approaches its orthogonal value $W_{\rm
  c}^{\rm O}$, see Fig. \ref{fig:phasediagram}. Also, for larger
values of $J$, it will stay at $W_{\rm c}^{\rm O}$ as a consequence of
the increase of Kondo screening with the exchange coupling $J$.
  
  \subsection{ Critical Semimetal Phase. }
 For
smaller exchange couplings $j<j^\ast$ a paradoxical situation
appears: The position of the critical point $W^I_{\rm c}$ depends on the
direction from which the AMIT is approached. Therefore, for
intermediate disorder strengths, $W_{\rm c}^{\rm O}< W < W^I_{\rm c}
(j)$ there exists a {\it critical region}. Accordingly, the mobility
edge is extended to a {\it critical band} whose width is a function of
$j$. The resulting zero-temperature quantum phase diagram is shown in
Fig. \ref{fig:phasediagram}. According to Eq. (\ref{wcj}) with Eq. (\ref{ts0}), the width of the 
 semimetal phase $W^I_{\rm c} (J) - W_{\rm c}^{\rm O} \sim 
 n_{M}^{\kappa(j)}$ increases with a power of the density of magnetic impurities $n_{M}$.

\section{Kondo-Anderson Transitions in a Magnetic Field}
\label{sec:magneticfield}

 A Zeeman magnetic field polarizes the free magnetic moments. 
  Thereby, their contribution to the 
   spin relaxation rate becomes diminished by the magnetic field.\cite{bobkov,vavilov,micklitz}
 On the other hand, the  Kondo singlet which Kondo screened magnetic moments form 
  with the conduction electrons 
  is  partially broken up by  the Zeeman  field. Thus, these magnetic moments 
     contribute a
      spin relaxation rate which is increasing 
       with the Zeeman field.  
      Finally, an orbital magnetic field breaks the time reversal symmetry
        and therefore also results in a   shift of the AMIT towards the unitary limit. 
   It is therefore an intriguing problem how  these  
         competing magnetic field effects combine to 
          change  the quantum phase diagram.\cite{raikh}

     The    magnetic field dependence of the spin relaxation rate from magnetic impurities
         in a metal at finite temperature was calculated in Ref. \onlinecite{vavilov}.
         The magnetic field polarises the magnetic impurity spins due to the Zeeman interaction, 
         \begin{equation} \label{zeeman}
 H_{Z} =-   \gamma _{s}  B \sum_{i} { S_{i z}, }
\end{equation}
where $  \gamma _{s} = g_{s} \mu_{\rm B} $ is the gyromagnetic ratio of the 
 magnetic impurities, $g_{s}$ their g-factor and $ \mu_{\rm B}$  is Bohr's magneton.
Here,  the magnetic field ${\bf B}$ is taken to point in the z-direction. 
         The   spin relaxation 
         rate from free magnetic impurities is found to be exponentially suppressed according to
          $1/\tau_{s} \sim \exp (-   \gamma _{s}  B |S_{z}|/k_{\rm B}T)$.
           Thus, in the zero temperature limit all free moments are expected to become polarized 
            by an arbitrarily small magnetic field, and their contribution to 
             the spin relaxation is  vanishing  identically.
             
\begin{figure}[htbp]
\begin{center}
\includegraphics[width=8.cm]{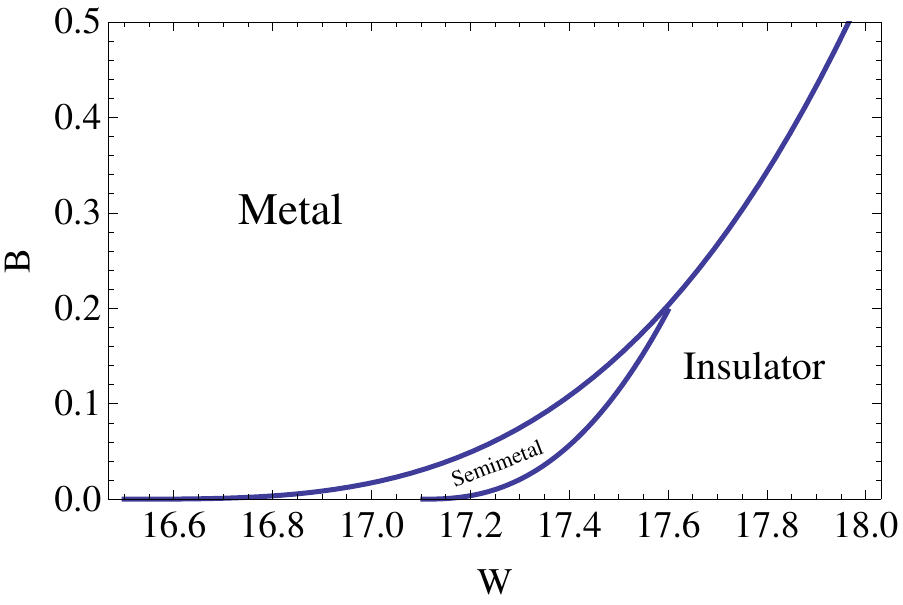}
\caption{(Color online) The quantum phase diagram in a magnetic field $B$ (arbitrary units),
 as function of disorder
 amplitude $W$, in units of $t$. We set $d=3,  \eta =2$ and $j=.2$. }   
\label{pdb}
\end{center}
\end{figure}

          Magnetic impurities which are screened
           with     a finite Kondo temperature $ T_{K}$, however, 
            have without magnetic field a  spin relaxation rate 
           Eq. (\ref{eq:tauapprox}) which vanishes in the low temperature limit.
             Applying a magnetic field the Kondo singlet is partially broken up 
              and a finite spin relaxation rate appears, which scales  with  $T_{K}$ 
            as\cite{vavilov,micklitz}
        \begin{equation} \label{srbtk}
\frac{1}{\tau_{ s}} (T_{K}) = \frac{n_{M}}{\pi \rho} \frac{(  \gamma_{s}  B |S_{z}|)^2}{T_{K}^2}.
\end{equation}
   To get the total spin relaxation rate, 
    we integrate over the distribution of Kondo temperatures $P(T_{K})$.
     We note  that the contribution from   magnetic impurities 
      with a small Kondo temperature, $T_{K} < g_{s} \mu_{\rm B} B |S_{z}|$
       vanishes since these spins become polarised. 
        The  ones with larger Kondo temperatures then yield  the spin relaxation rate, 
         \begin{equation} \label{srb}
\frac{1}{\tau_{ s}}  = \int_{  \gamma _{s} B |S_{z}|}^{\infty} d T_{K} P(T_{K}) 
 \frac{n_{M}}{\pi \rho} \frac{(  \gamma_{s} B |S_{z}|)^2}{T_{K}^2}.
\end{equation}
 For small magnetic fields, $  \gamma_{s} B |S_{z}| \ll T_{K}^{(0)} $,
  the main contribution comes from the low $T_{K}$-tail of the distribution, 
  \begin{equation}
P(T_K) \sim \left( \frac{E_c}{T_K} \right)^{1-j} \xi^{- \frac{d^2}{2  \eta }
  j^2 }.
\end{equation}
     Thus, we get 
              \begin{equation} \label{srb}
\frac{1}{\tau_{ s}}  =  
 \frac{n_{M}}{\pi \rho} \frac{d j}{2-j} \xi^{-\frac{d^2}{2   \eta } (j+  \eta/d )^2} \left(\frac{  \gamma _{s}  B |S_{z}|}{E_{c}}\right)^{ j}.
\end{equation}
  Setting $X_{s } = \xi^2/D_{e} \tau_{s} =1$, we find that the {\it Zeeman field }
    shifts  the critical disorder to
    \begin{equation} \label{wcb} 
    W_{c}(B) = W_{c}^I(j) + W_{c}^O c_{M} \left(\frac{  \gamma _{s}  B |S_{z}|}{E_{c}}\right)^{j \kappa(j)},
    \end{equation}
 where
 $1/\kappa(j) = 2 \nu(1- \frac{d^2}{4   \eta} (j+   \eta/d )^2
   )$ and 
  $c_{M} = \left(d j n_{M}/((2-j) \pi \rho D_{e})\right)^{ \kappa(j)}$.  
      Thus, the transition between critical semimetal  and insulator  is shifted 
       in a magnetic field according to Eq. (\ref{wcb}) as plotted in Fig. (\ref{pdb}).
   
   The {\it orbital  magnetic field} is known\cite{larkin} to shift $W_{c}$ to 
    \begin{equation}
    W_{c}(B)  = W_{c}^O + W_{c}^O ( \pi e B/h)^{1/(2\nu)}.
    \end{equation}
     This determines the transition line between metal and semimetal, 
      since, the Zeeman field   contributes a slower dependence on $B$, 
     coming from the metal side of the transition we find 
      $W_{c}(B)  = W_{c}^O + W_{c}^O  \left(\frac{  \gamma _{s}  B |S_{z}|}{E_{c}}\right)^{1/\nu}.$
    
  For the transition between semimetal and insulator, we can conclude that for 
  \begin{equation}
  j < j_{Z} =   \eta/d  \left(  2 \sqrt{d+1 +d^2/\eta }/d -1-2/d \right),
  \end{equation}
   the shift of $W_{c}$ is dominated by the Zeeman field over the orbital magnetic field. 
  For, $ d=3,   \eta/d  =2/3$ one finds, $j_{Z}= 0.185$, so that the Zeeman field 
   effect dominates for realistic values of exchange couplings $j$ as seen in Fig. (\ref{pdb}).
       
\section{Finite Temperature Properties}
\label{sec:nfl}

   We can  now proceed to calculate finite temperature properties. 
     To this end, we  first derive  the density of free magnetic moments at temperature $T$.
     It can be obtained by  integrating $P(T_{K})$ according to 
      \begin{equation}
      n_{FM} (T) = n_{M}\int_{0}^T d T_{K} P(T_{K}).
      \end{equation}
       At low temperatures, $T \ll T_{K}^0$  the free moments are
        determined by  the tail of the distribution. 
       
    \subsection{Insulator}
\begin{figure}[h]
\begin{center}
\includegraphics[width=8.cm]{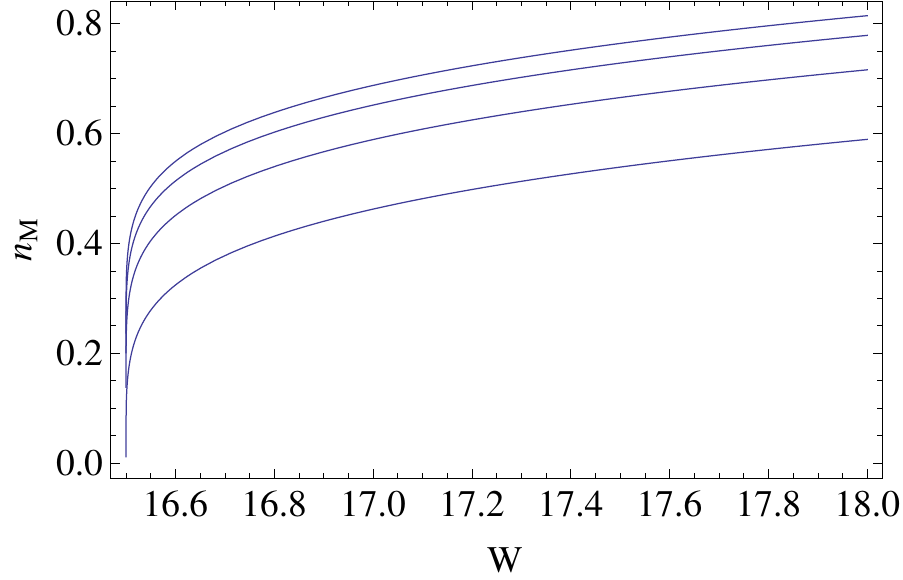}
\caption{(Color online) The ratio of free magnetic moments (magnetic moments with $T_{K}<T$), $n_{FM}/n_{M}$,
 as function of disorder strength $W$ on the insultating side of the transition
  for temperatures $T/E_{c}=0,1 \times 10^{-6},5 \times 10^{-6},1 \times 10^{-5}$ from the bottom to the top curve.
   We set $j=0.2$, $d=z=3$ and $  \eta/d  =2/3$.}   
\label{nfmt}
\end{center}
\end{figure}

       In the insulating regime for   $T >  \Delta_{\xi}$,  $P(T_{K})$ is  given by 
       Eq. (\ref{ptktail}),  while 
       at smaller temperature, not exceeding the local level spacing, 
       $T < \Delta_{\xi}$,   the tail of $P(T_{K})$  is changing to 
      Eq. (\ref{ptkxi}). Thus we find the density of free moments in the insulating regime 
            \begin{equation} \label{nfmti}
      n_{FM} (T) =n_{FM} (0) +  n_{M} \left\{
\begin{array}{cc}
  \frac{2d}{\eta} \left(\frac{T}{E_{c}} \right)^{\frac{\eta}{2d}} {\rm for} &  T >  \Delta_{\xi}  \\
    \frac{1}{j} \left(\frac{T}{E_{c}} \right)^{j \hspace{.3cm}}   \xi^{-\frac{
    1}{2   \eta } (d j)^2}{\rm for} & T <  \Delta_{\xi}
\end{array}
\right.
  ,
      \end{equation}
        where  the density of free moments at $T=0$,  $n_{FM} (0)$ is  given by Eq. (\ref{nfm}),
         decaying towards the AMIT. This result is plotted as function of disorder
          amplitude $W$ for various temperatures $T$ in Fig. (\ref{nfmt}).
          Thus, we find that the magnetic susceptibility 
           is diverging at low temperature 
                \begin{eqnarray}
         \chi (T) =\frac{n_{FM}(T)}{T} \sim n_{FM} (0)\frac{1}{T} +
         \nonumber \\
            n_{M}\frac{1}{E_{c}} \left\{ \begin{array}{cc}
  \frac{2d}{\eta} \left(\frac{T}{E_{c}} \right)^{\frac{\eta}{2d}-1} {\rm for} &  T >  \Delta_{\xi}  \\
    \frac{1}{j} \left(\frac{T}{E_{c}} \right)^{j-1 \hspace{.3cm}}  \xi^{-\frac{
    1}{2   \eta } (d j)^2}{\rm for} & T <  \Delta_{\xi}
\end{array}
\right. ,
         \end{eqnarray}
          with a Curie tail, whose weight increases as the Fermi energy moves deeper into the insulating regime. 
           The specific heat  is given by, 
            \begin{equation}
          C (T) \sim T \frac{d
  n_{\rm FM} (T)}{d T} = n_{M}   \left\{ \begin{array}{cc}
 \left(\frac{T}{E_{c}} \right)^{\frac{\eta}{2d}} {\rm for} &  T >  \Delta_{\xi}  \\
 \left(\frac{T}{E_{c}} \right)^{j \hspace{.3cm}}  \xi^{-\frac{
    1}{2   \eta } (d j)^2} {\rm for} & T <  \Delta_{\xi}
\end{array}
\right. .
  \end{equation}
    These results also apply at the AMIT, where $n_{FM} (0) \rightarrow 0$ and $\Delta_{\xi}
     \rightarrow \Delta$.

       \subsection{ Metal}   In the metallic regime, 
         $P(T_{K})$ is for $T < \Delta_{\xi}$ given by 
        Eq. (\ref{ptktailmetal}),
         so that we obtain,
         \begin{equation} \label{nfmtm}
      n_{FM} (T) = n_{M} \frac{d}{\eta} \left( \frac{T}{E_{c}} \right)^{\frac{\eta}{2d}} \exp \left[
      - \frac{\Delta_{\xi}^{\frac{\eta}{d}}}{c_{1}} \ln (\frac{T}{T_{K}^0})^2  \right], 
      \end{equation}      
        Accordingly, we find that the magnetic susceptibility $\chi (T) =n_{FM}(T)/T$ 
         has also in the metallic regime a 
 power law tail, which is however  cutoff at $T < \Delta_{\xi}$, where it converges 
  to zero.                  
We  get the contribution  to the specific heat, using $C (T) \sim T \frac{d
  n_{\rm FM} (T)}{d T}  \sim  n_{\rm FM} (T)$,
vanishing at low temperatures $T \ll  \Delta_{\xi}$. 

   One may ask if these  thermodynamic results are modified by inelastic 
    scatterings, introducing a 
    thermal length $L_{T}$,  which decreases with increasing temperature
    according to $L_{T} \sim T^{-1/z}$, where $z$ is the  dynamical exponent.
     If   electron-electron and electron-phonon scatterings are disregarded, 
      $L_{T}$ is set by the relation 
    $T = \Delta_{L_{T}}= D/L_{T}^d$,
     yielding $z=d$. 
      Therefore, it has been argued that  on length scales exceeding 
      $L_{T}$,  
      the system size $L$ and the localisation length/correlation lengths
      $\xi$ are substituted by $L_{T}$,  in the scaling theory of the AMIT \cite{leeramakrishnan}.        
          Thus, indeed, for temperatures $T > \Delta_{\xi}$,
           $\xi$ would have to be substituted by $L_{T}$. However, 
            as we find above, in this temperature range the
             results do not depend on $\xi$ anymore, so that 
              the finite $L_{T}$ does not modify the above results.

\section{Finite Temperature Phase Diagram: Kondo-Anderson Transitions} 
\label{sec:finiteT}

 The spin relaxation rate is found to   depend on
temperature,  Eq. (\ref{eq:tauapprox}), due to the Kondo screening, and the 
 temperature dependence of 
the density of free magnetic moments,  Eq. (\ref{nfmti}). Therefore,  the amount by which the 
 spin- and time-reversal-symmetry is broken,  depends on
temperature as well. Since the position of the AMIT is determined by  these
symmetries,\cite{ohtsukislevinreview}  it 
shifts as function of temperature $T$: when the spin relaxation rate
increases with temperature, the AMIT shifts towards 
 larger values of disorder amplitude
$W$. Thus, a metal-insulator
transition may occur at a finite temperature $T_{c}(W,J)$. In order to
 investigate the existence of such a transition, we will apply  the
Larkin-Khmel'nitskii condition \cite{larkin,droese}  with the temperature
dependent  symmetry parameter
$X_{s} (T) = \xi^2/D_{e} \tau_{s}(T)$.  Since the condition
$X_{s}=1$ gives an estimate for the  position of the
transition, $W_{c} (J,T)$, we find, 
\begin{equation}
\left[\xi(W_{c} (J,T)\right]^2 = a_{c}^2 \left[ \frac{W_{c}^{O}}{W_{c}
    (J,T)-W_{c}^{O}} \right]^{2\nu} = D_{e} \tau_{s}(J,T).
\end{equation}
Again, as in the previous section, we can apply this criterion in two ways:

\subsection{ Approaching the AMIT from the Insulator side.}
Coming from the insulating side of the transition, where   the
localization length $\xi$ is still finite and 
 smaller than the thermal length $L_{T}$, 
  the ratio
  $X = \xi^2/D_{e} \tau_{s}(T)$ is finite, giving a
measure of the amount of  time reversal symmetry breaking. $1/\tau_{s}(T)$ saturates
at low temperatures to the spin relaxation rate from free magnetic
moments Eq. (\ref{taufm}).  Thus, at low temperatures
 the transition occurs at $W_{c}^I(J)$ as given by Eq. (\ref{wcj}). At higher temperatures $1/\tau_{s}(T)$
increases with a power which depends on the
distribution of the Kondo temperature. 
  Thus,  the spin relaxation rate at finite temperature 
    is given by a weighted integral over the distribution function 
    of $T_{K}$, as $1/\tau_{s} (T) = \int_{0}^{\infty} d T_{K} P(T_{K}) 1/\tau_{s} (T/T_{K}) $.
  The spin relaxation rate of a magnetic impurity with a given Kondo temperature
   $T_{K}$, $ 1/\tau_{s} (T/T_{K})$,
   follows Eq. (\ref{eq:tauapprox}). Thus, it 
    increases first like $T^2/T_{K}^2$ 
    when $T < T_{K}$, until it 
reaches a maximum and then decays logarithmically slowly towards its
classical value $1/\tau^{\rm classical}_{s}$.  Since our analysis is
limited to low temperatures $T \ll T^0_{K}$, we can 
 simplify
$1/\tau_{s} (T)$ as a sum of spin relaxation of free moments of density Eq. (\ref{nfmti}),
  at sites whose density of states is suppressed as $\rho({\bf r})\sim \xi^{d-\alpha_{FM}}$,
   and the spin relaxation from spins whose Kondo temperature exceeds $T$,
    which we can approximate by $T^0_{K}$ due to the peaked distribution. 
     Thereby we get in good approximation the spin relaxation rate in the insulator  as 
     \begin{equation}
     \frac{1}{\tau_{s}} (T) = \frac{1}{\tau^0_{s}} \left( \xi^{d-\alpha_{0} - d j}\frac{n_{FM} (T)}{n_{M}}  + \frac{T^2}{T^{0 2}_{K}  } \frac{1- \frac{n_{FM} (T)}{n_{M}} }{S(S+1) \pi^2} \right).
     \end{equation}
      
\begin{figure}[t]
\includegraphics[width=8.2cm]{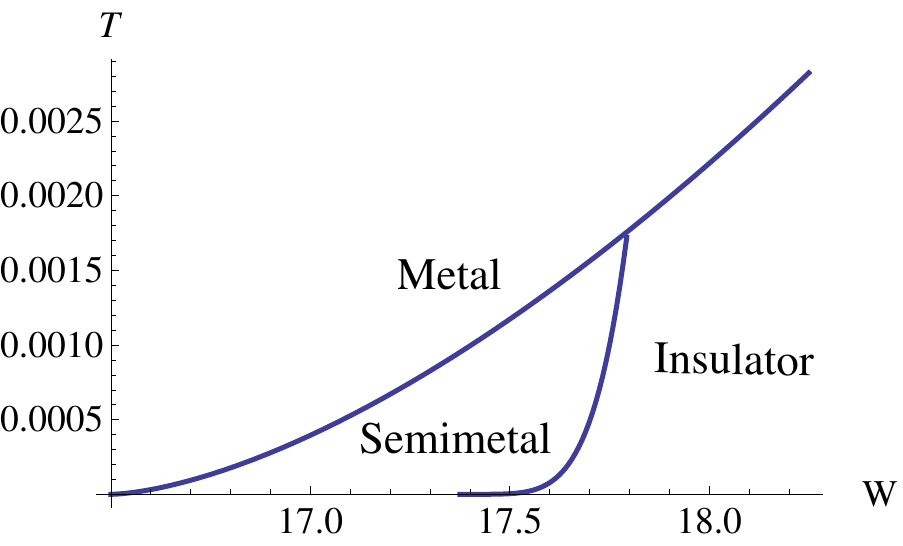}
\caption{(Color online) The finite-temperature phase diagram of
  Kondo-Anderson transitions. The solid lines are plots of the
  critical temperatures $T_{c}^M(W,J) $, Eq. (\ref{tcm}), and
  $T_{c}^I(W,J)$, Eq. (\ref{tci}), respectively, where the disorder
  amplitude $W$ is given in units of the hopping parameter $t$ and the
  temperature  is in ratios of  $E_{c}$. We used the following
  parameters: $j =0.2$, $\alpha_{0}=4$, $d=3$, $  \eta/d  =2/3$, $\nu=1.57$ and
  $a_{c}^2/(D_{e}\tau^{(0)}_{s}) = 0.1$. }
\label{fig:Tphasediagram}
\end{figure}

When $X_s=1$  the symmetry breaking is
sufficient to shift the transition to the larger disorder amplitude
$W_{c}^I(J,T)$.
 Accordingly, we find 
the  transition temperature,
\begin{equation} \label{tci}
T_{c}^I =E_{c } c_{I}\left(
\frac{W-W_{c}^{I}}{W_{c}^{O}} \right)^{\frac{1}{ j
      }},
\end{equation}
 where
 $W_{c}^{I}$ is given by Eq. (\ref{wcj}), and 
  $c_{I}=  (\kappa_{j}/j)^{-1/j} (D_{e} \tau_{s}^0)^{\kappa_{j}/j}$,
   where $1/\kappa_{j} = \nu (2-d^2/2/\eta (j+\eta/d)^2)$.

\subsection{ Approaching the AMIT from the Metallic side.}
Coming from the
metallic side, the density of free magnetic moments  is decaying fast 
 at  temperatures $T < \Delta_{\xi}$   according to  Eq. (\ref{nfmtm}).
  Thus,  the spin relaxation rate is dominated by the screened magnetic impurities, 
   yielding, 
\begin{equation} \label{tsfl}
\frac{1}{\tau_{s}} \approx \frac{1}{\tau^{0}_{s} } \frac{1}{S(S+1)\pi^2} \frac{T^2}{T_{K}^{0 2}},
\end{equation}
 where $1/\tau^{0}_{s}$ is given by Eq. (\ref{ts0}).
   Thus, the transition is shifted to
$W_{c}^M (T)$ and, accordingly, we find a transition temperature, 
\begin{equation} \label{tcm}
T_{c}^M (W) = \sqrt{ S(S+1) \pi^2 D_{e} \tau_{s}^0 } 
\left| \frac{W-W_{c}^{O} }{ W_{c}^{O} } \right|^{  \nu } T_{K}^{0},
\end{equation}
This  is plotted in Fig. \ref{fig:Tphasediagram} as function of disorder amplitude $W$.

Thus, we can conclude that there is a critical semimetal region which
extends over a finite temperature range, $T_{c}^I < T < T_{c}^M$.
  Since we derived the scaling function
only at small symmetry breaking parameter $X_s$,  the
phase diagram at larger disorder, where the critical disorder of the
unitary ensemble (when time-reversal symmetry completely broken) is
approached, might be modified. This part of the phase diagram is further complicated by
the fact that $1/\tau_{s}$ reaches a maximum and decays
logarithmically at temperatures exceeding $T^{(0)}_{K}$. Furthermore,
as inelastic scattering and dephasing processes will become stronger
at higher temperatures, that higher temperature part of the phase
boundaries is expected to become less well defined, mainly indicating
a crossover region.

\section{Conclusions and Discussion}
\label{sec:conclusions}

We conclude that spin correlations in disordered metals do strongly
modify the disorder induced Anderson metal-insulator
transitions. Starting from the numerically well-established fact that
a system of noninteracting electrons in a nonmagnetic disorder
potential undergoes a second-order transition between a metallic state
and an insulator,
we studied the mutual influence of Kondo correlations and the Anderson
localization transition. Since the position of the AMIT depends very
strongly on time-reversal and spin symmetries, we find that the
critical density and the critical disorder depend on the density of 
 magnetic impurities,  since these randomize the
conduction electron spins.  However, since the magnetic impurity
spin in a metal is screened at low temperatures due to the Kondo
effect, restoring a renormalized Fermi liquid with full time-reversal
symmetry,\cite{nozieresfl} we conclude that  coming from
the metallic side, the transition occurs at the critical disorder of a
time-reversal invariant system, $W_c^{\rm O}$. However,  coming from the
insulating side, it occurs instead at the stronger critical disorder
of a system where this symmetry is to some degree broken by
free magnetic 
 impurity spins.
 Since a
shift in the AMIT results in an exponential change of the finite
temperature resistivity, this has very strong experimental
consequences. Since both the correlation length and the localization
length are infinite in the critical region in-between the two critical
points, the zero-temperature conductivity vanishes in this
region, making the system a semimetal. 

Taking into account the multifractality of the critical wave
functions, we  derived the concentration of unscreened magnetic
impurities, the resulting spin scattering rate, and therefore the
shift of the AMIT.  Information on the multifractal distribution of
the critical wave functions, which is well established by numerous
careful numerical calculations,\cite{rmpmirlinevers} allowed us also
to characterize this critical region in more detail. Unscreened magnetic moments
 occur at sites with low intensity, which 
break the time-reversal symmetry of the conduction electrons. 
At  sites with high intensity local singlets
 are formed, where
one conduction electron is captured and localised, leaving the
symmetry of the other conduction electrons unchanged. Thus, the
Kondo-Anderson transitions share some features of a first-order phase
transition: In the critical semimetal phase there is coexistence of
 different magnetic states of the magnetic impurity spins, which
change the states of the conduction electrons, accordingly.
 A magnetic Zeeman fields is found to shift the transition to larger
  disorder $W$, and is dominating the magnetic field dependence
   for small exchange couplings $ j < j_{z}$.

An experimental consequence of the critical semimetal phase is the
divergence of the dielectric susceptibility, $\chi (T=0) \sim \xi^2
\sim [W^I_c(J)-W]^{-2\nu}$ at the disorder amplitude $W^I_c(J)$ (or at
a critical electron density $n^I_c(J)$), while the zero temperature
limit of the resistivity diverges with the correlation length as $\rho
(T=0) \sim \xi_c \sim (W-W_c^{\rm O})^{-\nu}$ at the weaker disorder
amplitude $W_c^{\rm O}$ (or at a larger critical density $n_c^{\rm
  O}$, accordingly). This difference, $W^I_c(J) - W_c^{\rm O}$,
increases with the concentration of Kondo impurities, as given by
Eq. (\ref{wcj}), until it converges to  its limiting value
  $W^U_c - W_c^{\rm O}$.\cite{ohtsukislevinreview} We
expect the critical semimetallic phase to be observable in materials where
both the AMIT and the Kondo effect are present simultaneously at
experimentally accessible temperatures, such as in amorphous
metal-semiconductor alloys \cite{kmck,amormetal} with dilute magnetic
impurities,\cite{hasegawa} or in doped semiconductors, such as Si:P,
where thermopower measurements are consistent with the presence of
Kondo impurities with $ \langle T_K \rangle \approx 1
K$.\cite{thermopower}

On the insulating side of the transition, physical properties such as
the finite-temperature resistivity are governed by the single-particle
gap $\Delta_{c} =1/(\rho \xi^d)= E_{c }[(W-W_{c}^I)/W_{c}^I]^{\nu d}$,
which causes activated behavior.
  
Aspects of the Kondo-Anderson transitions may be relevant for the
understanding of metal-insulator transitions in real materials like
Si:P. There have been detailed observations of a finite density of
magnetic moments in Si:P across the transition, from magnetic
susceptibility and specific heat measurements.\cite{loehneysen} These
have been previously modeled by the phenomenological two-fluid model
of Refs. \onlinecite{andres,paalanen} which did not take into account
the Kondo effect. On the other hand, a realistic description of Si:P
should consider that uncompensated Si:P contains a random array of
half-filled donor levels that are randomly coupled. At low-donor
concentration, close to the MIT, these levels are only weakly hybridized with the
conduction band. Therefore, Si:P should at least be described by a
half-filled Hubbard model with random hopping and on-site
potentials.\cite{bhatt,loehneysen, paalanen} Thus, all localized
states carry single electrons with spin $1/2$ and double occupancy is
prevented by the on-site interaction $U$. For the Hubbard model
without disorder, the ground state at half filling is
antiferromagnetic and there is an excitation gap of order
$U$. Dynamical mean-field theory (DMFT) calculations show that, at
least in the limit of infinite dimensions, spectral weight is
transferred into the middle of the gap; it has been argued that these
states yield extended quasiparticles which may form a Fermi
liquid.\cite{nozieres2004} These quasiparticles interact with the
localized magnetic moments in the lower Hubbard band,
 similarly to  normal conduction electron quasiparticles which interact with magnetic Anderson
impurities. In the presence of disorder, it has been shown that even
when there are no resonant levels, and the donor levels have merged
with the conduction band, the localized states in the band tail may
carry localized magnetic moments.\cite{milovanovic} Calculations of
the disordered Hubbard model, using the statistical
DMFT,\cite{Miranda,mirandaprl} and a variant of the local DMFT method
\cite{byczuk} show also that this model shares many features with the
disordered Anderson impurity model studied in this article. As
mentioned above, starting from the Anderson impurity model, we
neglected the magnetic impurity scalar potential $K$ given in
Eq. (\ref{knn}). We can justify this now, by referring to
nonperturbative numerical studies of the Kondo effect with a pseudogap
of power $r$ which find that, below the critical exchange coupling
$J_{c }= r D$, a scalar scattering potential is irrelevant and the
magnetic moment remains unscreened,\cite{pseudogap} leaving our
conclusion about the density of free moments unchanged. In the other
case of power-law divergencies, which we find to occur in the
multifractal states at sites with large intensiy ($\alpha <
\alpha_{0}$), a nonperturbative renormalization group analysis shows
that the on-site potential is irrelevant for all antiferromagnetic
exchange coupling amplitudes,\cite{bulla} leaving our conclusion on
the formation of local singlets at such sites unchanged as well.
   At finite density of magnetic moments the indirect exchange coupling 
    is in competition with the Kondo screening. In a disordered metal 
     these couplings are widely distributed. In the vicinity of the AMIT, 
      their distribution function can be calculated in a similar manner than 
       the Kondo temperature, and this will be presented in the subsequent article Ref. \onlinecite{rkkykondo}. 

A nonphenomenological, analytical approach to the AMIT of interacting,
disordered electrons is provided by field theoretical methods that
typically employ the replica trick.\cite{mitspinfluctuations}
Decoupling the resulting action of interacting Grassmann fields with
the Hubbard-Stratonovich transformation, a subsequent gradient
expansion of the matrix fields results in the nonlinear sigma
model,\cite{mitspinfluctuations,rmpbelitzkirkpatrick}. This kind of
field theory treats the disorder scattering nonperturbatively and
allows for a renormalization group analysis. It turns out that not
only the conductance but also a frequency parameter and the spin
density are renormalized. In particular, it has been found that the
disordered electron system becomes unstable to local spin density
fluctuations at the AMIT.\cite{mitspinfluctuations} It is still
unclear if these indicate the onset of a magnetic phase
transition,\cite{chamon00,nayak03} or are rather precursors of the
formation of local magnetic moments.\cite{rmpbelitzkirkpatrick} Since
the nonlinear sigma model is formulated in order to account for long
wavelength fluctuations such as charge diffusion and spin diffusion
modes, it cannot, at least in these early formulations,  describe
local moment formation. Therefore, it is an open challenge, to combine
the field-theoretical approach to the AMIT with a nonperturbative
treatment of the local moment formation and the Kondo correlations
caused by them. Field theoretical formulations of the disordered Kondo
problem have been developed in the mean field
approximation,\cite{burdin,kim} and in the Ginzburg-Landau
theory.\cite{kiselev,coqblin} However, in these approaches, the AMIT
of the conduction electrons in the disordered potential has not yet
been taken into account. We hope that the theory of the Kondo-Anderson
transitions presented in this paper, which was based, to some extent
phenomenologically, on the multifractal distributions and correlations
at AMITs, can serve as a guide in the quest for a  theory, which is 
derived from the Hamiltonian of disordered, interacting electrons,
taking into account nonperturbatively both disorder and interaction
effects, including Kondo correlations.

\section{Acknowledgements}

We gratefully acknowledge useful discussions with Elihu Abrahams, Ravin Bhatt, Georges
Bouzerar, Vladimir Dobrosavljevic, Alexander M. Finkel'stein,
Serge Florens, Hilbert von L\" ohneysen, Philippe Nozi\`eres, Mikhail Raikh, 
and Gergely Zarand. 
This research was supported by WCU(World Class University) program through the National Research Foundation of Korea funded by the Ministry of Education, Science and Technology(R31-2008-000-10059-0), Division of Advanced Materials Science. This work was partially financed by the Hungarian Research Fund (OTKA) grant
K73361 
and the Alexander von Humboldt Foundation.
 ERM, KS  and IV thank
the WCU AMS and APCTP for its hospitality. SK thanks the ILL Grenoble for its  hospitality.

\appendix

\section{Wave function Correlations in the vicinity of the AMIT}
\label{sec:A}

\subsection{ One Energy at the  mobility edge $E_{M}$.}

 The long-range spectral correlations in a $d$-dimensional system can
be quantified by spatially integrating the correlation function of the
eigenfunction probabilities associated to two energy levels distant by
$\omega_{nm} =E_n- E_m$.\cite{ioffe} When one of these energies is at the 
 mobility edge $E_{M}$, one
finds
\begin{eqnarray}
\label{criticalcorrelations}
 C_{nm} &=& L^d \int d^dr\, \left\langle |\psi_n({\bf r})|^2
 |\psi_m({\bf r})|^2 \right\rangle \nonumber \\ &=& \left\{
 \begin{array}{ll}(\frac{E_c}{{\rm Max} (|\omega_{nm}|, \Delta) })^ { \eta/d }, 
 & 0 < |\omega_{nm}| < E_c, \\ (E_c/|\omega_{nm}|)^{2}, &
  E_{p} > |\omega_{nm}| > E_c,
\end{array}
\right.
\end{eqnarray}
where $0<   \eta < d$, with $  \eta = 2 (\alpha_{0}-d)$.
  This exponent is obtained by the requirement that in the 
   limit of small energy differences, $|\omega_{nm}|  \rightarrow \Delta$,
    one recovers 
    \begin{equation}
C_{nm} |_{|\omega_{nm}|  \rightarrow \Delta} \rightarrow L^{2 d}\left\langle |\psi_n({\bf r})|^4
  \right\rangle  \sim L^{d  -\tau_{2}},
 \end{equation}
 where $\tau_{2}= d_{2} = d-2 (\alpha_{0}-d)$ and $\Delta = D/L^d$.
 For
$|\omega_{nm}| < E_c$, correlations are enhanced in comparison to the
plane-wave limit, where $C_{nm}=1$. Note that for $|\omega_{nm}| > E_c$
 the correlation function decays below $1$. This anticorrelation 
  ensures that the intensity is normalised: A dip in the intensity 
   at one energy implies an enhancement of intensity at another energy 
    in the band. This anticorrelation is expected to occur up to some 
     finite energy $E_{p}$, beyond which the correlation function 
     increases to the uncorrelated value $1$ in a nonuniversal way. 
     
\subsection{ Both energies at a distance from the  mobility edge $E_{M}$. }   

  The spectral correlations in a $d$-dimensional system also exist away
from the transition whenever either the correlation length $\xi$ (on
the metallic side) or the localization length $\xi_{c}$ (on the
insulator side) are finite. In the correlation function of
Eq. (\ref{criticalcorrelations}), the energy difference
$|\omega_{nm}|$ is for $\epsilon_{n} < \epsilon_{m}$ then substituted by ${\rm Max} [|\omega_{nm}|,  
\Delta_{\xi_{n}}]$, where $\Delta_{\xi_{n}} = E_{c}\,
(\xi_{n}/a_{c} )^{-d}$, and $a_{c}$ is defined by $E_{c}=1/(\rho
a_{c}^d)$, with $\rho$ denoting the average density of states. We get therefore 
\begin{eqnarray}
\label{criticalcorrelations2}
 C_{nm} &=& L^d \int d^dr\, \left\langle |\psi_n({\bf r})|^2
 |\psi_m({\bf r})|^2 \right\rangle \nonumber \\ &=& \left\{
 \begin{array}{ll} \left[ \frac{E_c}{{\rm Max} [|\omega_{nm}|,  
 \Delta_{\xi_{n}}] }\right]^{  \eta/d} , & 0 <
   |\omega_{nm}| < E_c, \\ \left( \frac{E_c}{|\omega_{nm}|}
     \right)^{2}, & |\omega_{nm}| > E_c,
\end{array}
\right.
\end{eqnarray}
where  $  \eta  = 2 (\alpha_{0}-d)$. For
$\omega_{nm} < E_c$, correlations are still enhanced in comparison to
the plane-wave limit, where $C_{nm}=1$.

\subsection{ Joint Distribution Function.}

The joint distribution function of two eigenfunction intensities at
the same position should be log-normal  since the distribution
function of a single eigenfunction intensity is log-normal as given by Eq. (\ref{eq:Pone}).
 Therefore, we recently
conjectured it  recently to be of a log-normal form, 
when one of the energies is at the mobility edge.\cite{kettemann}
In general, 
 when, one or both energies are away from the mobility edge, the finite correlation 
 length/ localisation length $\xi$ needs to be taken into account when one or both energies
  are on the metallic/localised side of the transition, respectively. 
We can  then conjecture the  joint distribution function for
$\alpha_{l}$ and $\alpha_{k}$, where, in the metallic regime, we
define
$
\alpha = - \frac{\ln [ (L/\xi)^d |\psi(x)|^2]}{\ln \xi},
$
which has the distribution $P(\alpha) \sim
\xi^{-(\alpha-\alpha_{0})^2/(2  \eta  )}$.
Accordingly, in the localized regime, 
 we get the same distribution function, 
  defining there $\alpha = - \ln [|\psi(x)|^2]/\ln \xi$. 
 Then, for
$|\epsilon_{l}-\epsilon_{k}| < E_{c}$, we
conjecture  the joint distribution function to be of the form, 
\begin{equation}
\label{eq:Ptwom}
P(\alpha_l,\alpha_k) = \xi_{l}^{ a_{l k} \left[ f (\alpha_l) -d
    \right]} \xi_{k}^{ a_{l k} \left[ f (\alpha_k) -d \right]} K_{l
  k}^{- a_{l k} \frac{(\alpha_l - \alpha_0)(\alpha_k -\alpha_0) }{d
      \eta }},
\end{equation}
where $\xi_{l}$ is the correlation/localization  length of a state at energy
$E_{l}$, $K_{{lk}} = {\rm Max} \left[ | E_{l} -E_{k} |, {\rm Min}(
  \Delta_{\xi_{l}}, \Delta_{\xi_{k}}) \right]/E_{c} $, and
\begin{equation}
a_{l k} = \frac{1}{1- \left( \frac{\ln K_{lk}}{d \ln \hat{\xi}}
  \right)^2 },
\end{equation}
where we introduced the length scale $\hat{\xi}$ through
\begin{equation}
\ln \hat{\xi} = \sqrt{\ln \xi_{l} \ln \xi_{k}}.
\end{equation}
Averaging the local intensity of the state with energy $E_{l}$ with
the conditional probability $P(\alpha,\alpha_{l})/P(\alpha)$, 
which can be rewritten as 
\begin{equation}
\label{conditional}
P_{\alpha_k=\alpha} (\alpha_l) = \xi_{l}^{- a_{lk}\frac{[\alpha_l - \alpha_0 +
      \frac{\ln K_{lk}}{d \ln \xi_{l}} (\alpha - \alpha_0)]^2}{2   \eta  }},
\end{equation}
we then
get Eq. (\ref{intensitymetallic}),
\begin{equation}
I_{\alpha {\rm }} (\xi_{l},\xi_{k})= \langle L^d |\psi_{l}
({\bf r})|^2 \rangle_{\alpha} = K_{lk}^{\frac{\alpha-\alpha_{0}}{d}
  -\frac{  \eta }{2 d^2} \frac{ \ln K_{l k} }{ \ln \xi_{k}} }.
\end{equation}

   Note that when the energy $E_{k}$ is  at the 
      mobility edge $E_{k }=E_{M}$  then 
      $\xi_{k} \rightarrow L$, 
       in Eq. (\ref{eq:Ptwom}),  and we 
      recover the conditional intensity at energy $E_{l}$, Eq. (\ref{cci}), in a more
       rigourous way (the finite correlation length of that state, $\xi_{l}$, did not appear
        explicitly  in  
        our previous conjecture Eq. (7) of Ref. \onlinecite{kettemann}, but is now taken into account in 
         Eq.  (\ref{eq:Ptwom})).
         
         \subsection{ Higher Moment Correlation Functions}
        
      The correlation of  higher moments of the intensities is defined by 
         \begin{eqnarray}
\label{criticalcorrelationsq}
 C^q_{\{n_{i}\}} &=& L^{q d} \left\langle \prod_{i=1}^q |\psi_{n_{i}}({\bf r})|^2
\right\rangle. \end{eqnarray}
 We consider first the case, where, one of the energies is fixed to the 
  mobility edge, $E_{n_{1}}= E_{M}$.
  The power law of the correlation of  higher moments of the intensity, 
  can then be obtained by taking the limit of small energy differences, 
  $\omega_{n_{i}, n_{j} } \rightarrow \Delta$,
   which yields, 
     \begin{eqnarray}
\label{criticalcorrelationsqq2}
 C^q_{\{ n_{i}\}} \mid_{\omega_{n_{i}, n_{j}} \rightarrow \Delta, \forall i}&=& L^{q d} \left\langle  |\psi_{n}({\bf r})|^{2 q} \right\rangle 
 \nonumber \\
 & \sim & L^{(q -1)d -\tau_{q}} \sim L^{\eta_{q}}.
 \end{eqnarray}
$\tau_{q}$ is found to  terminate for $q > q_{c} = \alpha_{0}/\eta$.
  Thus, for $d=3$ and $\alpha_{0}=4$, $q_{c} =2$, so that all higher order moments
   have the same  $\tau_{q}= \tau_{2} = d-\eta, \forall q > q_{c}$.
    Thus we find 
    \begin{equation}
    \eta_{q} = (q-2)d  + \eta.
    \end{equation}
  Furthermore, we notice that the pair of 
  intensities whose energy are closest to each other are correlated with 
   a power $\eta/d$.
   Thus,   we can conclude that  the 3-rd order correlation function 
   is to leading order given by 
\begin{eqnarray} \label{c3}
C^3_{ \{ n_{i}\}}  \sim
  \left(\frac{E_c}{|{\rm Min} (\omega_{n_{i},n_{j} }) | }\right)^{ \frac{\eta}{d} } 
\frac{E_c}{|{\rm Max} ( \omega_{n_{i},n_{j}} )| }.
\end{eqnarray}
 This result is valid for  $   0 < |\omega_{n_{i},n_{j}}| < E_c   \forall   i,j$.
 If all energies in $C^3$ are placed away from the  mobility edge, $E_{n_{1}} \neq E_{M}$,
  then  the energy difference 
   $|{\rm Min} ( \omega_{n_{i},n_{j}})|$, is replaced by $\Delta_{n_{i}} = E_{c}/\xi_{n_{i}}^d$ if 
     it 
   is smaller than that energy scale, respectively.

\section{$F[\alpha,T_{K}]$}
\label{sec:B}
  Let us first consider the case, when the Fermi energy is at the mobility edge. 
     We can then evaluate $F[\alpha,T_{K}]$ in the following way. 
      We transform the 
      summation over states $l$ to the 
      integration over energy $\epsilon$. Then, we transform  to  
      $ t = - \ln (\epsilon/E_{c})$.
        Next, we can substitute in good approximation 
         $\tanh x \approx  x$ for $0< x < 1$, and $\tanh x \approx  1$ for $ x > 1$.
     Thereby, we get 
     \begin{eqnarray} \label{f1}
     F[\alpha,T_{K}] =  j \left( \int_{\ln (E_{c}/T_{K})}^{\ln (E_{c}/\Delta)} dt e^{-t} + 
      \int^{\ln (E_{c}/T_{K})}_{0} dt \right)
      \nonumber \\
      \times  \exp [ - \frac{\alpha - \alpha_{0}}{d} t - \frac{\alpha_{0}-d}{d^2}
      \frac{t^2}{\ln L}  ].
     \end{eqnarray}
      The integrals can now be performed, yielding Error functions,
       $\int_{a}^b e^{-A t^2 - B t} = e^{B^2/(4A)} \frac{\sqrt{\pi}}{2 \sqrt{A}} 
        (Erf [\sqrt{A} b+ B/(2\sqrt{A}))] - Erf[\sqrt{A} a+ B/(2\sqrt{A}))] )$, for each term 
         in Eq. (\ref{f1}). Since $A= (\alpha_{0}-d)/(d^2 \ln L)$ the arguments of 
          all Error functions diverge for $L \rightarrow \infty$, 
           and we can use the asymptotic expansion $Erf[z]  \rightarrow {\rm sign}
            z \times 1  - e^{-z^2}/(\sqrt{\pi} z)$.
             Thereby, we find for $\alpha > \alpha_{0}-d$,
             \begin{eqnarray} \label{ftk}
              F[\alpha,T_{K}]/j
              & =  & \frac{d}{\alpha -\alpha_{0}}   + \left( \frac{d}{2 \alpha - \eta} - \frac{d}{\alpha -\alpha_{0}} \right)  \left( \frac{E_{c}}{T_{K}} \right)^{\frac{\alpha_{0}-\alpha}{d}} 
              \nonumber \\ 
                & - & d_{\alpha} \frac{E_{c}}{T_{K}} L^{-\alpha},
             \end{eqnarray}
             where $d_{\alpha} = d/(\eta +2 \alpha)$.
                Solving $1 = F[\alpha,T_{K}]$  we obatin then for $L \rightarrow \infty$, Eq. (\ref{tkalpha})
                 for $T_{K}(\alpha)$.\\

%





\begin{thebibliography}{11}

 \bibitem{bhatt} R. N. Bhatt and D. S. Fisher, Phys. Rev. Lett. {\bf
   68}, 3072 (1992).

\bibitem{langenfeld} A. Langenfeld and P. W\" olfle, Ann. Physik {\bf
  4}, 43 (1995).

\bibitem{dobros} V. Dobrosavljevic, T. R. Kirkpatrick, and
  G. Kotliar, Phys. Rev. Lett. {\bf 69}, 1113 (1992); E. Miranda,
  V. Dobrosavljevic, and G.  Kotliar, {\it ibid.} {\bf 78}, 290
  (1997).

\bibitem{Miranda} E. Miranda and V. Dobrosavljevic,
  Rep. Prog. Phys. {\bf 68}, 2337 (2005).

\bibitem{lernerldos} I. V. Lerner, Phys. Lett. A {\bf 133}, 253
  (1988).

\bibitem{kettemannjetp} S. Kettemann and E. R. Mucciolo, Pisma
  Zh. Eksp. Teor.  Fiz. {\bf 83}, 284 (2006) [JETP Lett. {\bf 83}, 240
    ( 2006)]; S. Kettemann and E. R. Mucciolo, Phys. Rev. B {\bf 75},
  184407 (2007)

\bibitem{micklitz} T. Micklitz, A. Altland, T. A. Costi, and A. Rosch,
  Phys. Rev. Lett. {\bf 96}, 226601 (2006); T. Micklitz, T. A. Costi, and A. Rosch,
  Phys. Rev. B {\bf 75}, 054406 (2007).
  
  

\bibitem{cornaglia} P. S. Cornaglia, D. R. Grempel, and
  C. A. Balseiro, Phys. Rev. Lett. {\bf 96}, 117209 (2006).

\bibitem{imre} I. Varga, E. R. Mucciolo, and S. Kettemann, 
 International Journal of Modern Physics: Conference Series,   Vol. x, to be published 
  (2012).
  
  
\bibitem{zhuravlev} A. Zhuravlev, I. Zharekeshev, E. Gorelov,
  A. I. Lichtenstein, E. R. Mucciolo, and S. Kettemann,
  Phys. Rev. Lett. {\bf 99}, 247202 (2007).

\bibitem{multifractal} F. Wegner, Z. Phys. B {\bf 36}, 209 (1980);
  H. Aoki, J. Phys. C {\bf 16}, L205 (1983); C. Castellani and
  L. Peliti, J. Phys. A {\bf 19}, L991 (1986); M. Schreiber and
  H. Gru\ss bach, Phys. Rev.  Lett. {\bf 67}, 607 (1991); M. Janssen,
  Int. J. Mod. Phys. B {\bf 8}, 943 (1994).

\bibitem{rmpmirlinevers} F. Evers and A. D. Mirlin,
  Rev. Mod. Phys. {\bf 80}, 1355 (2008).

\bibitem{rodriguez} L. J. Vasquez, A. Rodriguez, and R. A. Romer,
  Phys. Rev. B {\bf 78}, 195106 (2008).

\bibitem{RKKY} M. A. Ruderman and C. Kittel, Phys. Rev. {\bf 96}, 99
  (1954); T. Kasuya, Prog. Theor. Phys. {\bf 16}, 45 (1956);
  K. Yosida, Phys. Rev. {\bf 106}, 893 (1957).
  
\bibitem{liechtenstein} A. I. Liechtenstein, M.  I. Katsnelson,
  V. P. Antropov, and V.  A. Gubanov, J. Magn. Magn. Mater. {\bf 67},
  65 (1987).
  
  \bibitem{rkkykondo} S. Kettemann, K. Slevin, E. Mucciolo, unpublished (2011). 
%


\bibitem{powerlaw} J. T. Chalker, Physica A (Amsterdam) {\bf 167}, 253
  (1990); V. E. Kravtsov and K. A. Muttalib, Phys. Rev. Lett. {\bf
  79}, 1913 (1997); J. T. Chalker {\it et al.}, JETP Lett. {\bf 64},
  386 (1996); T. Brandes, B. Huckestein, L. Schweiser,
  Ann. Phys. (Leipzig) {\bf 5}, 633 (1996); V. E. Kravtsov, {\it
    ibid.} {\bf 8}, 621 (1999); V. E. Kravtsov, A. Ossipov,
  O. M. Yevtushenko, and E. Cuevas, Phys. Rev. B {\bf 82}, 161102
  (2010); V. E.  Kravtsov, A.  Ossipov, O. M.  Yevtushenko,
J. Phys.  {\bf A 44},    305003   (2011). 

\bibitem{cuevas} E. Cuevas and V. E. Kravtsov, Phys. Rev. B {\bf 76},
  235119 (2007).

\bibitem{ioffe} M. V. Feigel'man, L. B. Ioffe, V. E. Kravtsov, and
  E. A. Yuzbashyan, Phys. Rev. Lett. {\bf 98}, 027001 (2007).

\bibitem{ioffe2} M. V. Feigel'man, L. B. Ioffe, V. E. Kravtsov,
  E. Cuevas, Annals of Physics {\bf 325}, 1368 (2010).

\bibitem{kettemann} S. Kettemann, E. R. Mucciolo, and I. Varga,
  Phys. Rev. Lett. {\bf 103}, 126401 (2009).
  
  
  \bibitem{mirlin} A. D. Mirlin, Phys. Rep. {\bf 326}, 259 (2000).
\bibitem{andersonmm} P. W. Anderson, Phys. Rev. {\bf 124}, 41(1961).

\bibitem{hewson} A. C. Hewson, {\it The Kondo Problem to Heavy
  Fermions} (Cambridge University Press, Cambridge, 1997).
 
\bibitem{pseudogap} D. Withoff and E. Fradkin, Phys. Rev. Lett. {\bf
  64}, 1835 (1990); K. Ingersent, Phys. Rev. B {\bf 54}, 11936 (1996);
S. Florens, M. Vojta,   Phys. Rev. B {\bf 72}, 115117 (2005).
  L. Fritz, S. Florens, and M. Vojta, Phys. Rev. B {\bf 74}, 144410
  (2006).
  

\bibitem{suhl} Y. Nagaoka, Phys. Rev. {\bf 138}, 1112 (1965); H. Suhl,
  Phys. Rev. {\bf A 138}, 515 (1965).
  

\bibitem{zu96} G. Zarand and L. Udvardi, Phys. Rev. B {\bf 54}, 7606
  (1996).
  
  \bibitem{remark}  For simplicity, we 
    restrict here  the integration interval to $E_{c} = D/(d \ln 2d)$. 
     Since the integration should actually be extended upto the band edges, 
      the Kondo temperature we obtain is accordingly smaller  
      by a constant factor. Since we rescale  all results with the 
      Kondo temperature  of a clean system, $T_{K}^0$,  as obtained in the same approximation, 
      $T_{K}^0 = E_{c} \exp (+1/2 -1/j )$ this 
       approximation does not effect any of the results reported here. 
       

%
%



\bibitem{bulla} M. Vojta, and R. Bulla, Eur. Phys. J. B {\bf 28}, 283
  (2002).

  
  
  	\bibitem{kmkprl}
A. MacKinnon and B. Kramer,
Phys. Rev. Lett. 47, 1546 (1981); A. MacKinnon, and B. Kramer,
 Zeitschrift f. Physik {\bf B 53},  1 (1983).


\bibitem{meraikh} S. Kettemann and M. E. Raikh, Phys. Rev. Lett. {\bf
  90}, 146601 (2003).
  

\bibitem{larkin} D. Khmelnitskii and A. I. Larkin, Solid State
  Comm. {\bf 39}, 1069 (1981).

\bibitem{droese} T. Dr\" ose, M. Batsch, I. Kh. Zharekeshev, and
  B. Kramer, Phys. Rev. B {\bf 57}, 37 (1998).
  
 \bibitem{wegner1986} 
  F. J. Wegner, Nuclear Physics B270 [FS16] , 1(1986).
  
  
\bibitem{hikami} S. Hikami, A. I. Larkin, and Y. Nagaoka,
  Prog. Theor. Phys. {\bf 63}, 707 (1980).

  
  \bibitem{dielectric}
  M. A. Paalanen, T. F. Rosenbaum, G. A. Thomas, and R. N. Bhatt, 
  Phys. Rev. Lett. {\bf 51}, 1896 (1983).
  
  \bibitem{leeramakrishnan} P. A. Lee and T. V. Ramakrishnan, Rev. Mod. Phys. {\bf 57}, 287 (1985).
  
\bibitem{bergmann} G. Bergmann, Phys. Rev. Lett. {\bf 58}, 1236
  (1987); R. P. Peters, G. Bergmann, and R. M. Mueller, {\it ibid.}
  {\bf 58}, 1964 (1987); C. Van Haesendonck, J. Vranken, and
  Y. Bruynseraede, {\it ibid.} {\bf 58}, 1968 (1987).

\bibitem{mw00} P. Mohanty and R. A. Webb, Phys. Rev. Lett. {\bf 84},
  4481 (2000).

\bibitem{zarand} G. Zar\'and, L. Borda, J. von Delft, and N. Andrei,
  Phys. Rev. Lett. {\bf 93}, 107204 (2004).
  
  
\bibitem{nozieresfl} P.  Nozi\`eres, J. Low Temp. Phys. {\bf 17}, 3
  (1974).
  
  
\bibitem{metalspinrelaxation} A finite magnetic
scattering rate may arise on the metallic side when clusters of $k$
magnetic impurities form. However, their probability and the resulting
magnetic scattering rate scale with the $k$th power of the density of
magnetic impurities,\cite{boucai} and therefore are negligible for
dilute systems.


\bibitem{boucai} E. Boucai, B. Lecoanet, J. Pilon, J. L. Tholence, and
  R. Tournier, Phys. Rev. B {\bf 3}, 3834 (1971).
  
\bibitem{ohtsukislevinreview} T. Ohtsuki and T. Kawarabayashi,
  J. Phys. Soc. Jpn. {\bf 66}, 314 (1997); T. Ohtsuki, K. Slevin, and
  T. Kawarabayashi, Ann. Physik {\bf 8}, 655 (1999).

\bibitem{roemer} A. Rodriguez, L. J. Vasquez, K. Slevin, R. A. Roemer,
 Phys. Rev. B 84, 134209 (2011).

  
\bibitem{bobkov}  A.A. Bobkov, V.I. FalÕko, and D.E. KhmelÕnitskii, Zh. Exp.
Teor. Fiz. {\bf 98}, 703 (1990) [Sov. Phys. JETP {\bf 71}, 393
(1990)].

  \bibitem{vavilov} M.G. Vavilov, L.I. Glazman,
  Phys. Rev. B 67, 115310 (2003);
  M. G. Vavilov, L. I. Glazman, and A. I. Larkin, 
Phys. Rev. B 68, 075119 (2003).
%


\bibitem{raikh} We thank M. Raikh for asking us this question. 
  



\bibitem{andres} K. Andres, R. N. Bhatt, P. Goalwin, T. M. Rice, and
  R. E. Walstedt, Phys. Rev. B {\bf 24}, 244 (1981) .


  
  
\bibitem{footnote1}
Note that we could get a more accurate result
  by integrating Eq. (\ref{eq:tauapprox}) $T_{K}$, weighted by the
  distribution of $T_{K}$, Eq. (\ref{ptkxi}) in the localized
  regime. The result for low temperatures will not be strongly changed
  from the following result, however.

\bibitem{amormetal} C. Van Haesendonck and Y.  Bruynseraede,
  Phys. Rev. B {\bf 33}, 1684 (1986); B. W. Dodson, W. L. McMillan,
  J. M. Mochel, and R. C. Dynes, Phys. Rev. Lett. {\bf 46}, 46 (1981).

\bibitem{kmck} B. Kramer, A. MacKinnon, Rep. Prog. Phys. {\bf 56},
  1469 (1993).
  

\bibitem{hasegawa} C. C. Tsuei and R. Hasegawa, Solid State
  Commun. {\bf 7}, 1581 (1969).

\bibitem{thermopower} M. Lakner and H. v. L\"ohneysen,
  Phys. Rev. Lett. {\bf 70}, 3475 (1993).

\bibitem{loehneysen}
 H. v. L\"ohneysen, Adv. in Solid State Phys. {\bf 40}, 143 (2000).
 
 \bibitem{finkelsteinesr} A. M. Finkel'shtein, JETP Lett. {\bf 46},
 513 (1987). 
 
 
 \bibitem{esr} M. A. Paalanen, S. Sachdev,  R. N. Bhatt, and
  A. E. Ruckenstein, Phys. Rev. Lett. {\bf 57}, 2061 (1986).

\bibitem{paalanen} M. A. Paalanen, J. E. Graebner, R. N. Bhatt, and
  S. Sachdev, Phys. Rev. Lett. {\bf 61}, 597 (1988); S. Sachdev,
  Phys. Rev. B {\bf 39}, 5297 (1989).
  
\bibitem{nozieres2004} P. Nozieres, J. Stat. Phys.  {\bf 115}, 19
  (2004).
  
\bibitem{milovanovic} M. Milovanovic, S. Sachdev, and R. N. Bhatt,
  Phys. Rev. Lett. {\bf 63}, 82 (1989).
  
\bibitem{mirandaprl} E. Miranda and V. Dobrosavljevic,
  Phys. Rev. Lett. {\bf 86}, 264 (2001); M. C. O. Aguiar,
  V. Dobrosavljevic, E. Abrahams, and G.  Kotliar,
  Phys. Rev. Lett. {\bf 102}, 156402 (2009).
  
  
\bibitem{sato} K. Sato, L. Bergqvist, J. Kudrnovsky, P. H. Dederichs,
  O. Eriksson, I. Turek, B. Sanyal, G. Bouzerar, H. Katayama-Yoshida,
  V. A. Dinh, T. Fukushima, H. Kizaki and R. Zeller,
  Rev. Mod. Phys. {\bf 82}, 1633 (2010).

\bibitem{byczuk} K. Byczuk, W. Hofstetter, U. Yu, and D. Vollhardt,
  Eur. Phys. J. Special Topics {\bf 180}, 135 (2010).
  
\bibitem{mitspinfluctuations} A. M. Finkel'stein, JETP Lett. {\bf
  37}, 517 (1983); {\it ibid.} {\bf 40}, 796 (1984).
 
\bibitem{rmpbelitzkirkpatrick} D. Belitz and T. R. Kirkpatrick,
  Rev. Mod. Phys. {\bf 66}, 261 (1994).

\bibitem{chamon00} C. Chamon and E. R. Mucciolo, Phys. Rev. Lett. {\bf
  85}, 5607 (2000).

\bibitem{nayak03} C. Nayak and X. Yang Phys. Rev. B {\bf 68}, 104423
  (2003).

\bibitem{burdin} S. Burdin, and P. Fulde, Phys. Rev. B {\bf 76},
  104425 (2007).

\bibitem{kim} M.-T. Tran and K.-S. Kim, Phys. Rev. Lett. {\bf 105},
  116403 (2010);   J. Phys. Condens. Matter 23, 425602 (2011).

\bibitem{kiselev} M. N. Kiselev, K. Kikoin, and R. Oppermann,
  Phys. Rev. B {\bf 65}, 184410 (2002).


\bibitem{coqblin} S. G. Magalhaes, F. M. Zimmer, B. Coqblin,
  Phys. Rev. B {\bf 81}, 094424 (2010); B. Coqblin, J. R. Iglesias,
  N. B. Perkins, S. G. Magalhaes, F. M. Zimmer, J. Magn. Magn. Mater.
  {\bf 320}, 1989 (2008).
  
\end{thebibliography}
\end{document}